\documentclass[preprint,12pt]{elsarticle}

\usepackage{etex}
\usepackage{longtable}
\usepackage{colortbl}
\usepackage{examplep}
\usepackage{arydshln}
\usepackage[colorlinks, hypertexnames=false]{hyperref}
\AtBeginDocument{
  \hypersetup{
    linkcolor=red,
    citecolor=blue,
  }
}
\usepackage{amssymb}
\usepackage{graphicx}
\usepackage{ctable}
\usepackage{fancyhdr}
\usepackage{footnote}
\usepackage{cleveref}
\usepackage{fancyvrb}
\usepackage{placeins}
\usepackage{array}
\usepackage{lipsum}
\usepackage{pstricks,pstricks-add}
\usepackage{array,multirow}
\usepackage{setspace}
\usepackage{amsthm}
\usepackage{bbold}
\usepackage{scrextend}
\usepackage{hyperref}
\usepackage{units}
\usepackage[font=small,format=plain,labelfont=bf,up,textfont=normal,up]{caption}
\usepackage{amssymb}
\usepackage{amsmath}
\usepackage{footmisc}
\usepackage{blkarray}
\usepackage[margin=3cm]{geometry}
\usepackage[
singlelinecheck=false,format=hang 
]{caption}
\usepackage{setspace}
\numberwithin{equation}{section}

\newcommand{\bigcell}[2]{\begin{tabular}{@{}#1@{}}#2\end{tabular}}

\newcounter{bla}
\journal{Computer Physics Communications}

\begin{document}

\begin{frontmatter}



\title{Implementation of the manifest left-right symmetric model in FeynRules}


\author[a]{Aviad Roitgrund\corref{author}}
\author[a]{Gad Eilam}
\author[a]{Shaouly Bar-Shalom}

\cortext[author] {Corresponding author.\\\textit{E-mail address:} aviadroi@tx.technion.ac.il}
\address[a]{Technion-Israel Institute of Technology, 32000 Haifa, ISRAEL}
 \vspace{0.9cm}
\begin{abstract}
We present an implementation of the manifest left-right symmetric model in FeynRules. The different aspects of the model are briefly described alongside the corresponding elements of the model file. The model file is validated and can be easily translated to matrix element generators such as \verb+MadGraph5_aMC@NLO+, \verb+CalcHEP+, Sherpa, etc. The implementation of the left-right symmetric model is a useful step for studying new physics signals with the data generated at the LHC.
\end{abstract}


\end{frontmatter}



\section{introducton}
\label{sec.introduction}
The main goal of the LHC is to search for signals of new physics beyond the Standard Model (SM),  motivated
 by the shortcomings of the SM. In particular, the SM is incapable of explaining a number of fundamental issues, such as the hierarchy problem (resulting from the large difference between the
 weak force and the gravitational force), dark matter, the number of families in the quark and lepton sector.
 It is, therefore, widely believed that new physics beyond the SM will be discovered in the coming years. Among the possible attractive platforms for new physics are left-right symmetric models (LRSM)\cite{LRSM1,LRSM2}, on which we focus in this paper.
In particular, we describe here a LRSM with no explicit CP violation in the Higgs potential, the manifest (or quasi-manifest) left-right symmetric model (MLRSM or QMLRSM, respectively), and its implementation in matrix element generators through \verb+FeynRules 2.0+ \cite{feynrules}.

The LRSM address two specific difficulties of the SM: (i) Parity violation in the weak interactions, and (ii) non-zero neutrino masses implied by the experimental evidence of neutrino oscillation \cite{massive neutrinos}. In particular, the left-right symmetry which underlies LRSM restores Parity symmetry at energies appreciably higher than the electroweak (EW) scale, resulting in the addition of three new heavy gauge bosons, $W_2^{\pm}$ and $Z_2$. Furthermore, in LRSM the neutrinos are massive, where their nature (i.e., whether they are of Majorana or Dirac type) depends on the details of the LRSM.

Early constructions of the LRSM comprise a Higgs sector with a Higgs bidoublet and two Higgs doublets \cite{LRSM1}.
In such a setup, the neutrinos are of Dirac type and no natural explanation for their small
masses is provided. A later version, the above mentioned MLRSM, incorporates a Higgs bidoublet and two Higgs triplets, which leads to Majorana type neutrinos \cite{LRSM2}. In particular, the MLRSM provides a natural setup for the smallness of neutrino masses, relating their mass scale to the large left-right symmetry breaking scale through the see-saw mechanism \cite{classic}.

This work includes the following:
\begin{itemize}
\item
An implementation of the MLRSM carrying an identical structure to the Lagrangian of \cite{gluza} (i.e. identical parametrization and definitions of the Lagrangian terms). As such, it features the following elements in particular:
\begin{enumerate}
\item
The use of alternative empirical parameters (i.e. fermion mass matrices, CKM-type mixing matrices and Higgs VEVs) to indirectly set values to the Yukawa matrices,
\item
Majorana type neutrinos (for a Dirac type see the Left-Right model based on \cite{previous}), 
\item
Directly controlling the QMLRSM diagonal matrices described in Refs.\cite{classic} and \cite{gluza}.
\end{enumerate}
\item
As a result, this is the first MLRSM implementation which includes a fully verified Lagrangian (by comparison to Refs.\cite{classic} and \cite{gluza}) as part of a thorough validation procedure performed according to the guidelines in Refs.\cite{feynrules, guidelines} (see below).
\end{itemize}

The model file created in this work was generated using the \verb+FeynRules 2.0+ package \cite{feynrules}
(running on \verb+Mathematica 7.0+ \cite{mathematica}). It contains a methodically built model Lagrangian with a-priory defined ingredients (i.e. the underlying symmetries, gauge fields, mixing angles and Higgs fields), and its output is the computed Feynman rules of the MLRSM/QMLRSM. The user can easily translate the model to a selection of matrix element generators. The model file was thoroughly validated and uploaded to URL \newline \url{https://drive.google.com/folderview?id=0BxMAGX_Tlpi9X0RUZW9tS2RaQ0E&usp=sharing}.

The paper is organized as follows. In section \ref{sec.General model description} we describe the LRSM Lagrangian field content and structure. In section \ref{sec.The Higgs sector} we present the Higgs sector, and describe
 the corresponding spontaneous symmetry breaking (SSB) mechanism in this model. As part of the discussion about the Higgs sector we also include:
\begin{itemize}
\item
A brief summary of the constraints stemming from the minimization conditions of the Higgs potential and from the explicit CP conservation requirement.
\item
A presentation of the gauge eigensystem which emanate from the kinetic terms of the Higgs Lagrangian.
\item
A description of the Yukawa sector and how to obtain its Feynman rules in physical basis as mentioned above.
\end{itemize}
We continue in section \ref{sec.implement} with a description of the model implementation, and explain the different aspects of the model file, such as the field definitions,\linebreak the mixing and Yukawa matrices. We then explain how the user can control the different parameters in the model. In section \ref{sec.Validation} we validate the model file using the \verb+FeynRules 2.0+ interface and the matrix element generators \verb+CalcHEP v.3.6.23+ \cite{calchep} and \verb+MadGraph5_aMC@NLO v.2.2.3+ \cite{madgraph}. 
Finally, in section \ref{sec.Conclusion}, we summarize. The list of particles and parameters of the model along with the corresponding model file notations is given in \ref{App:AppendixA}. This list is followed by the list of the user controlled parameters of the model file. These two lists are connected through the expressions given in \ref{App:AppendixB} which relate the model parameters to the user controlled parameters. In \ref{appendix.Higgs physical eigenstates} we present the Higgs multiplet fields in terms of the physical Higgs eigenstates, and in \ref{appendix.comparsion-settings} we list the parameter values chosen for the validation procedure.

\section{General model description}
\label{sec.General model description}
\subsection{The field content of the model}
\label{sec.The field content of the model}

The LRSM is based on the gauge group $SU(3)_C \times SU(2)_L \times SU(2)_R \times U(1)_{B-L}$. All the fermion fields in the model are assigned to doublets, including the right handed fermions which transform as doublets under the new symmetry of the model, $SU(2)_R$. In addition to the fermion fields, seven gauge fields - $\vec{W}_{L,R}$ and $B$ (corresponding to the groups $SU(2)_{L,R}$ and $U(1)_{b-L}$)
and  eight gluon fields - $G_a\,(a=1..8)$ are introduced in order to obtain gauge invariance.
The scalar content of the model includes three Higgs multiplets: a bidoulbet (denoted as $\phi$), a right handed and a left handed triplet (denoted as $\boldsymbol{\Delta}_L$ and $\boldsymbol{\Delta}_R$, respectively). The covariant derivatives of the multiplets are conventionally given in the adjoint representation, so that the triplets are also converted to the adjoint representation. Thus,
the $2 \times 2$ bidoublet-equivalent field matrices $\Delta_{L,R}=\frac{1}{\sqrt{2}}\vec{\sigma}\cdot\boldsymbol{\Delta}_{L,R}$ are introduced (the three-vector $\vec{\sigma}$ contains the Pauli matrices as components). The model field content is given in Table \ref{tab:table1}.
\begin{longtable}{| l  l  l  l  l  l |}
    \hline \hline
\rowcolor[gray]{0.85}\rule{0pt}{4ex} Model fields & content & $SU(3)_C \quad \times $ & $SU(2)_L \quad\times$ & $SU(2)_R \quad\times$ & $U(1)_{B-L}$ \\
    \hline
\endfirsthead
\multicolumn{6}{c}%
{{\bfseries \tablename\ \thetable{} -- continued from previous page}} \\
\hline \hline
\rowcolor[gray]{0.85}\rule{0pt}{4ex} Model fields & content & $SU(3)_C \quad \times $ & $SU(2)_L \quad\times$ & $SU(2)_R \quad\times$ & $U(1)_{B-L}$ \\
\hline
\\
\endhead
\hline
\multicolumn{6}{|r|}{{Continued on next page}} \\ \hline
\endfoot
\endlastfoot
\rule{0pt}{4ex}
    \hspace {0.35cm} $L_{iL}$ & $\left( \begin{array}{c} \nu'_i \\
    l'_i\end{array} \right)_L$ & \quad 1 & \quad 2 & \quad 1 & \quad -1 \\[15pt]
    \hspace {0.5cm} $L_{iR}$ & $\left( \begin{array}{c} \nu'_i \\
    l'_i\end{array} \right)_R$ & \quad 1 & \quad 1 & \quad 2 & \quad -1 \\[15pt]
    \hspace {0.5cm} $Q_{iL}$ & $\left( \begin{array}{c} u'_i \\
    d'_i\end{array} \right)_L$ & \quad 3 & \quad 2 & \quad 1 & \quad \,$\frac{1}{3}$ \\[15pt]
    \hspace {0.5cm} $Q_{iR}$ & $\left( \begin{array}{c} u'_i \\
    d'_i\end{array} \right)_R$ & \quad 3 & \quad 1 & \quad 2 & \quad \,$\frac{1}{3}$ \\[15pt]
\hline
\rule{0pt}{4ex}
    \hspace {0.5cm} $W_L$ & $W^+_L,\,W^-_L,\,W^3_L$ & \quad 1 & \quad 3 & \quad 1 & \quad 0 \\
\rule{0pt}{4ex}
    \hspace {0.5cm} $W_R$ & $W^+_R,\,W^-_R,\,W^3_R$ & \quad 1 & \quad 1 & \quad 3 & \quad 0 \\
\rule{0pt}{4ex}
    \hspace {0.5cm} $B$ & $\quad \quad B$ & \quad 1 & \quad 1 & \quad 1 & \quad 0 \\[5pt]
\hline
\\
\hspace {0.7cm} $\phi$ & $\displaystyle{
\left( \begin{array}{cc}
\phi^{0}_{1} & \phi^{+}_{1}\\
\phi^{-}_{2} & \phi^{0}_{2} \end{array} \right)}$ & \quad 1 & \quad 2 & \quad 2 & \quad 0 \\[15pt]
\hspace {0.7cm} $\Delta_R$ & $\left( \begin{array}{cc}
\frac{\delta^{+}_R}{\sqrt{2}} & \delta^{++}_R\\
\delta^{0}_R & -\frac{\delta^{+}_R}{\sqrt{2}} \end{array} \right)$  & \quad 1 & \quad 3 & \quad 1 & \quad 2 \\[20pt]
\hspace {0.7cm} $\Delta_L$ & $\left( \begin{array}{cc}
\frac{\delta^{+}_L}{\sqrt{2}} & \delta^{++}_L\\
\delta^{0}_L & -\frac{\delta^{+}_L}{\sqrt{2}} \end{array} \right)$  & \quad 1 & \quad 1 & \quad 3 & \quad 2 \\[20pt]
    \hline
\caption{The field content in the LRSM and the corresponding quantum numbers. The index $i=1,2,3$ of the fermionic fields runs over the number of generations. In addition, the $'$ in the fermion fields denotes  that these are gauge eigenstates.}
\label{tab:table1}
\end{longtable}

\subsection{The Lagrangian structure}
\label{subsec.Lag-struc}
The Lagrangian of the MLRSM can be divided into four terms:
\begin{align}
\mathcal{L}=\mathcal{L}_{kinetic}+\mathcal{L}_{gauge}+\mathcal{L}_{Yukawa}+\mathcal{L}_{Higgs}.
\label{eq.lagrangian}
\end{align}
The $\mathcal{L}_{kinetic}$ part contains the interactions between fermions and gauge bosons which are invariant under $SU(3)_C \times SU(2)_L \times SU(2)_R \times U(1)_{B-L}$. In particular, the fermionic kinetic terms take the following form
\begin{align}
L_{f}&=i\sum\bar{\psi}\gamma^\mu D_\mu\psi \nonumber \\
&=\bar{L}_L\gamma^{\mu}\left(i\partial_{\mu}+g_L\frac{\vec{\sigma}}{2}\cdot\vec{W}_{L\mu}-\frac{g'}{2}B_{\mu}\right)L_L \nonumber \\
&+\bar{L}_R\gamma^{\mu}\left(i\partial_{\mu}+g_R\frac{\vec{\sigma}}{2}\cdot\vec{W}_{R\mu}-\frac{g'}{2}B_{\mu}\right)L_R \nonumber \\
&+\bar{Q}_L^\alpha\gamma^{\mu}\Big[\left(i\partial_{\mu}+g_L\frac{\vec{\sigma}}{2}\cdot\vec{W}_{L\mu}+\frac{g'}{6}B_{\mu}\right)\delta_{\alpha\beta} + \frac{g_s}{2}\lambda_{\alpha\beta}\cdot G_\mu\Big]Q_L^\beta \nonumber \\
&+\bar{Q}_R^\alpha\gamma^{\mu}\Big[\left(i\partial_{\mu}+g_R\frac{\vec{\sigma}}{2}\cdot\vec{W}_{R\mu}+\frac{g'}{6}B_{\mu}\right)\delta_{\alpha\beta} +\frac{g_s}{2}\lambda_{\alpha\beta}\cdot G_\mu\Big]Q_R^\beta ~,
\label{lag.fermions}
\end{align}
The appropriate coupling constants of the  $G^a_\mu$, $\vec{W}_{L,R \,\mu}$ and the $B_\mu$ fields are $g_s$, $g_{L,R}$ and $g^\prime=g_{B-L}$, respectively. The requirement that the Lagrangian is invariant under the left-right symmetry
\begin{align}
\psi_L \leftrightarrow \psi_R,\quad \vec{W}_L \leftrightarrow \vec{W}_R,
\label{eq.lr-symmetry}
\end{align}
leads to
\begin{align}
g_L=g_R.
\end{align}
The gauge bosons kinetic terms and inner interactions are
\begin{align}
L_{gauge}=-\frac{1}{4}G^{\mu\nu}_a G_{a\mu\nu}-\frac{1}{4}W^{\mu\nu}_{Li}W_{Li\mu\nu}-\frac{1}{4}W^{\mu\nu}_{Ri}W_{Ri\mu\nu}-\frac{1}{4}B^{\mu\nu}B_{\mu\nu}, \label{gaugegauge}
\end{align}
where $G^a_{\mu\nu}$, $W_{L,R\mu\nu}^{i}$ and $B_{\mu\nu}$ are the field strength tensors of the $SU(3)_C$, $SU(2)_{L,R}$ gauge fields and the $U(1)_{B-L}$ gauge field, respectively. They are defined as follows:
\begin{align}
&G_a^{\mu\nu}=\partial^\mu G_a^\nu-\partial^\nu G_a^\mu-g_sf^{abc}G_b^\mu G_c^\nu & (a,b,c=1..8) \nonumber \\[4pt]
&W_{iL}^{\mu\nu}=\partial^\mu W^\nu_{Li}-\partial^\nu W^{\mu}_{Li}
+g_L \, \varepsilon^{ijk} \, W^\mu_{Lj}W^\nu_{Lk} & (i,j,k=1..3) \nonumber \\[4pt]
&W_{iR}^{\mu\nu}=\partial^\mu W^\nu_{Ri}-\partial^\nu W^{\mu}_{Ri}
+g_R \, \varepsilon^{ijk} \, W^\mu_{Rj}W^\nu_{Rk} & (i,j,k=1..3) \nonumber \\[4pt]
&B^{\mu\nu}=\partial^\mu B^\nu-\partial^\nu B^\mu.
\end{align}
where $f^{abc}$ and $\varepsilon^{ijk}$ are the structure constants of the $SU(3)_C$ and $SU(2)$ groups, respectively.

The Yukawa interactions part, $L_{Yukawa}$, consists of the most general possible couplings of the Higgs multiplets to bilinear fermion field products which form singlets under $SU(2)_L\times SU(2)_R \times U(1)_{B-L}$:
\begin{align}
\mathcal{L}_Y=-\sum_{i,j}\big(&\bar{L}_{iL}((h_L)_{ij}\phi+(\tilde{h_L})_{ij}\tilde{\phi})L_{jR} - \bar{Q}_{iL}((h_Q)_{ij}\phi+(\tilde{h_Q})_{ij}\tilde{\phi})Q_{jR} \nonumber \\
& -\overline{\strut{(L_{iR})}^c}\;\Sigma_R {(h_M)}_{ij} L_{jR} -\overline{\strut{(L_{iL})}^c}\;\Sigma_L {(h_M)}_{ij}L_{jL}\big)
+h.c.
\label{eq.yukawa}
\end{align}
where $\tilde{\phi}\,\equiv\, \tau_2\phi^*\tau_2$, $\Sigma_{L,R}=i\tau_2\Delta_{L,R}$ and $h_Q$, $h_L$,$h_M$,$\tilde{h_Q}$, $\tilde{h_L}$ are $3\times 3$ Yukawa matrices in flavor space. We will extract expressions for the Yukawa matrices in terms of empirical parameters (i.e. CKM type mixing matrices, diagonal mass matrices and Higgs VEVs) in sec. \ref{sub.yukawa}.

The Higgs Lagrangian term consists of the Higgs kinetic terms and the potential of the Higgs multiplets:
\begin{align}
\mathcal{L}_{Higgs}=\underbrace{\Sigma_i[Tr| D_\mu \Theta_i |^2]}_{\text{\color{blue}kinnetic terms}}-\underbrace{V_{Higgs}}_{\text{\color{red}potential}}
\label{eq.scalarL}
\end{align}
where $\Theta_i=\{\phi,\Delta_L,\Delta_R\}$. As mentioned above, the covariant derivatives for the Higgs multiplets are given in the adjoint representation. Under the $SU(2)_L \times SU(2)_R \times U(1)_{B-L}$ symmetry they are given by
\begin{align}
& D_{\mu}\phi =\partial_\mu\phi-i\frac{g_L}{2}\left(\vec{\sigma}\cdot\vec{W}_{L\mu}\right)\phi+i\frac{g_R}{2}\phi\left(\vec{\sigma}\cdot\vec{W}_{R\mu}\right), \nonumber \\
& D_{\mu}\Delta_{L,R}=\partial_{\mu}\Delta_{L,R}-i\,\frac{g_{L,R}}{2} \, \vec{W}_{L,R\mu} \cdot \left[\vec{\sigma}, \Delta_{L,R}\right]-ig'B_{\mu}\Delta_{L,R}.
\label{eq.covariant-derivatives}
\end{align}
After spontaneous symmetry breaking (SSB) in the Higgs sector, the charged and neutral gauge bosons acquire masses through the kinnetic terms in Eq.\eqref{eq.scalarL}. We present the gauge boson mass spectrum in sec.\ref{sub.eigensystem} as part of the discussion about the Higgs sector to which we shall now proceed.

\section{The Higgs sector}
\label{sec.The Higgs sector}
\subsection{Electroweak symmetry breaking and the Higgs multiplets}
The Higgs sector of the LRSM consists of multiplets which, as spontaneous left-right symmetry breaking occurs, acquire VEVs which contain the $U(1)_Q$ observed symmetry. The electromagnetic charge $Q$ is defined by the modified Gell-Mann-Nishijima formula
\begin{align}
Q=I_{3L}+I_{3R}+\frac{B-L}{2}.
\end{align}

Although the LRSM Lagrangian is invariant under $SU(2)_L \times SU(2)_R \times U(1)_{B-L}$, the vacuum states are not. These states acquire VEVs in a pattern which should be compatible with phenomenological requirements. To begin with, the symmetry breaking is required to be (partly) left-right symmetric by itself,
in order to obtain Dirac mass terms for the fermions.
This is accomplished by introducing the left-right symmetric Higgs bidoublet
(the numbers in the parenthesis stand for its properties under $SU(3)_C$, $SU(2)_L$ and $SU(2)_R$
and for its quantum number under $B-L$, respectively):
\begin{align}
\displaystyle{
\phi=\left( \begin{array}{cc}
\phi^{0}_{1} & \phi^{+}_{1}\\
\phi^{-}_{2} & \phi^{0}_{2} \end{array} \right)} \hspace{10 mm} (1,2,2,0) ~.
\end{align}
This bidoublet acquires VEVs for the neutral fields and generates the Dirac masses of the fermions
via its couplings to the fermion bilinears $\bar{f}_Lf_R$ and $\bar{f}_Rf_L$ ($f=Q,L$). This by itself is not yet sufficient to break the left-right symmetry of Eq.\eqref{eq.lr-symmetry} to $U(1)_Q$.\footnote{The reason is that the $B-L$ quantum number attributed to the bidoublet is zero and, in addition, the fields which acquire the VEVs have no electric charge. Therefore, after symmetry breaking, the vacuum remain symmetric under $U(1)_{B-L}\times U(1)_Q$ instead of just $U(1)_Q$ (see also \cite{mohapatra}).}
 Thus, as mentioned in the introduction, the model outlined here also employs (for the additional symmetry breaking) the two Higgs triplets:
\begin{align}
\Delta_{L,R}=\left( \begin{array}{cc}
\delta^{+}_{L,R}/\sqrt{2} & \delta^{++}_{L,R}\\
\delta^{0}_{L,R} & -\delta^{+}_{L,R}/\sqrt{2} \end{array} \right) \hspace{5 mm}\text{L}:(1,3,1,2) \hspace{3 mm} \text{R}:(1,1,3,2)
\end{align}
(the most general form of potential which consists of the above Higgs multiplets and obeys the left-right symmetry is shown in sec. \ref{sec.potential}).

The above setup is sufficient in order to further break the symmetry to $U(1)_Q$ in a two-stage process.
In the first stage, which takes place at an energy scale much larger than the electroweak scale, parity breaks down and the right handed Higgs triplet, $\Delta_R$, acquires a VEV $v_R$:
\begin{align}
<\Delta_R>=\frac{1}{\sqrt{2}}\left( \begin{array}{cc}
0&0\\ v_R &0\end{array} \right).
\label{eq.vevR}
\end{align}
The VEV of $\Delta_R$ violates the $B-L$ symmetry as its quantum number is chosen to be $B-L=2$:
\begin{align}
(B-L)<\Delta_R>=2<\Delta_R> \neq 0.
\end{align}
The initial LR symmetry is thus spontaneously broken, and one is left with the electroweak symmetry:
\begin{align}
SU(2)_L\,\otimes \, SU(2)_R \, \otimes \,  U(1)_{B-L} \xrightarrow{\textcolor{magenta}{<\Delta_R>}} SU(2)_L\,\otimes\,U(1)_Y
\label{eq.Symbreak1}
\end{align}
(generating masses for the heavy gauge bosons $W_2^\pm$ and $Z_2$, which are approximately proportional to $v_R$, see sec.\ref{sub.eigensystem}). 
The still unbroken hypercharge ($Y$) symmetry is defined as
\begin{align}
Y<\Delta_R>&=(2\,I_{3R}+(B-L))<\Delta_R>=\left(\begin{array}{cc} 0 & 0 \\
-v_R\sqrt{2} & 0 \end{array}\right) + \left(\begin{array}{cc} 0 & 0 \\
v_R\sqrt{2} & 0 \end{array}\right)=0.
\end{align}
At the second stage the bidoublet acquires a VEV
\begin{align}
<\phi>=\frac{1}{\sqrt{2}}\left( \begin{array}{cc}
k_1 & 0\\ 0& k_2
\end{array} \right),
\label{eq.vevbi}
\end{align}
which violates the above $Y$ symmetry due to its chosen $SU(2)_R$ and $B-L$ dimensions,
\begin{align}
Y<\phi>&=(2I_{3R}+(B-L))<\phi>=\sqrt{2}\left(\begin{array}{cc}-k_1 & 0 \\
0 & k_2 \end{array}\right)\neq 0,
\end{align}
but the electric charge symmetry still remains unbroken, since for example
\begin{align}
\hat{Q}<\phi>&=\big[I_3,<\phi>\big]=0.
\end{align}
The VEVs $k_1$ and $k_2$ give rise to (and are proportional to the masses of) the SM gauge bosons $W_L$ and $Z$.
Comparing these VEVs to $v_R$, which gives rise to the new, heavier and yet undiscovered gauge bosons $W_R$ and $Z_2$\footnote{\label{note0}This assumption is supported by theoretical and experimental lower limits on the heavy gauge boson masses. For example, the phenomenological constraint $W_R>\unit[1.6]{TeV}$ which arises from evaluating the $K_L-K_S$ mass difference from $\overline{K^0}-K^0$ mixing\cite{beall}. In addition, direct searches imply $W_R>\unit[1.8]{TeV}$, $Z'>\unit[1.3]{TeV}$\cite{pdg}.} (and, as mentioned above, is proportional to their masses), implies that
\begin{equation}
v_R \, \gg \, k_1,\, k_2.
\end{equation}

In addition to the above VEVs, also the left handed triplet $\Delta_L$ can acquire a VEV as a result of spontaneous symmetry breaking, given by
\begin{align}
<\Delta_L>=\frac{1}{\sqrt{2}}\left( \begin{array}{cc}
0&0\\ v_L&0\end{array} \right)~,
\label{eq.vevL}
\end{align}
which tends to zero, due to phenomenological considerations (see below, Eq.\eqref{eq.VEVseesaw-rem}). The second stage of the symmetry breaking sums up to
\begin{align}
SU(2)_L\,\otimes\,U(1)_Y \xrightarrow{\textcolor{magenta}{<\phi>,\,<\Delta_L>}} U(1)_Q
\end{align}
where the SM $W_L^\pm$ and $Z$ bosons acquire their masses, which are proportional to $k_1$ and $k_2$ when $v_L\rightarrow 0$ (see sec.\ref{sub.eigensystem}).

\subsection{The Higgs potential}
\label{sec.potential}
The most general scalar potential which is invariant under the left-right symmetry of the Higgs multiplets (see e.g., \cite{classic2}):
\begin{align}
\Delta_L\;\leftrightarrow\;\Delta_R,\quad\quad\phi\;\leftrightarrow\;\phi^\dagger,
\label{eq.scalarsym}
\end{align}
is
\allowdisplaybreaks[1]
\begin{align}
& V(\phi,\Delta_L,\Delta_R)= \nonumber \\&-\mu_{1}^2\left(Tr[\phi^\dagger\phi]\right)-\mu_{2}^2\left(Tr[\tilde{\phi}\phi^\dagger]
+\left(Tr[\tilde{\phi}^\dagger\phi]\right)\right)-\mu_{3}^2\left(Tr[\Delta_L\Delta_L^{\dagger}]
+Tr[\Delta_R\Delta_R^{\dagger}]\right) \nonumber \\&
+\lambda_1\left(\left(Tr[\phi\phi^\dagger]\right)^2\right)
+\lambda_2\left(\left(Tr[\tilde{\phi}\phi^\dagger]\right)^2
+\left(Tr[\tilde{\phi}^\dagger\phi]\right)^2\right)
+\lambda_3\left(Tr[\tilde{\phi}\phi^\dagger]Tr[\tilde{\phi}^\dagger\phi]\right)\nonumber\\&
+\lambda_4\left(Tr[\phi\phi^{\dagger}]\left(Tr[\tilde{\phi}\phi^\dagger]
+Tr[\tilde{\phi}^\dagger\phi]\right)\right) \nonumber \\
& +\rho_1\left(\left(Tr[\Delta_L\Delta_L^{\dagger}]\right)^2
+\left(Tr[\Delta_R\Delta_R^{\dagger}]\right)^2\right)\nonumber\\&
+\rho_2\left(Tr[\Delta_L\Delta_L]Tr[\Delta_L^{\dagger}\Delta_L^{\dagger}]
+Tr[\Delta_R\Delta_R]Tr[\Delta_R^{\dagger}\Delta_R^{\dagger}]\right) \nonumber \\
& +\rho_3\left(Tr[\Delta_L\Delta_L^{\dagger}]Tr[\Delta_R\Delta_R^{\dagger}]
\right)\nonumber \\
&+\rho_4\left(Tr[\Delta_L\Delta_L]Tr[\Delta_R^{\dagger}\Delta_R^{\dagger}]
+Tr[\Delta_L^\dagger \Delta_L^\dagger ]Tr[\Delta_R\Delta_R]\right)
\nonumber\\&
+\alpha_1\left(Tr[\phi\phi^{\dagger}]\left(Tr[\Delta_L\Delta_L^{\dagger}]
+Tr[\Delta_R\Delta_R^{\dagger}]\right)\right) \nonumber \\
&+\alpha_2\left(Tr[\phi\tilde{\phi}^{\dagger}]Tr[\Delta_R\Delta_R^{\dagger}]
+Tr[\phi^{\dagger}\tilde{\phi}]Tr[\Delta_L\Delta_L^{\dagger}]\right)
\nonumber \\
&+\alpha_2^{*}\left(Tr[\phi^{\dagger}\tilde{\phi}]Tr[\Delta_R\Delta_R^{\dagger}
]+Tr[\tilde{\phi}^{\dagger}\phi]Tr[\Delta_L\Delta_L^{\dagger}]\right) \nonumber \\
& +\alpha_3\left(Tr[\phi\phi^{\dagger}\Delta_L\Delta_L^{\dagger}]
+Tr[\phi^{\dagger}\phi\Delta_R\Delta_R^{\dagger}]\right)\nonumber\\&
+\beta_1\left(Tr[\phi\Delta_R\phi^{\dagger}\Delta_L^{\dagger}]
+Tr[\phi^{\dagger}\Delta_L\phi\Delta_R^{\dagger}]\right)
+\beta_2\left(Tr[\tilde{\phi}\Delta_R\phi^{\dagger}\Delta_L^{\dagger}]
+Tr[\tilde{\phi}^{\dagger}\Delta_L\phi\Delta_R^{\dagger}]\right)\nonumber\\&
+\beta_3\left(Tr[\phi\Delta_R\tilde{\phi}^{\dagger}\Delta_L^{\dagger}]
+Tr[\phi^{\dagger}\Delta_L\tilde{\phi}\Delta_R^{\dagger}]\right),
\label{eq.potential}
\end{align}
where $\mu_i$ are mass parameters and $\lambda_i$, $\rho_i$, $\alpha_i$, $\beta_i$ are dimensionless couplings.
With the exception of $\alpha_2$, all the parameters in the potential are real due to the left right symmetry of Eq.\eqref{eq.scalarsym}. Since in this work we assume that CP is explicitly conserved in the potential (in both the MLRSM and the QMLRSM) we also set $\alpha_2$ to be real (see \cite{gluza}).

As discussed in \cite{classic}, two of the VEVs phases, typically the phases of $v_R$ and of $k_1$, can be absorbed by global phase transformations. Thus there are six minimization conditions of the neutral fields, given by
\begin{align}
\frac{\partial V}{\partial v_R}=\frac{\partial V}{\partial k_1}=\frac{\partial V}{\partial\mathfrak{Re}k_2}=\frac{\partial V}{\partial\mathfrak{Re}v_L}=\frac{\partial V}{\partial\mathfrak{Im}k_2}=\frac{\partial V}{\partial\mathfrak{Im}v_L}=0.
\end{align}
The first three conditions can be used to solve the mass parameters squared, $\mu^2_1$, $\mu^2_2$ and $\mu^2_3$, in terms of the other parameters (see also \cite{kiers}). The requirement of explicit CP conservation in addition to the second three conditions implies that in addition to $v_R$ and $k_1$, the rest of the Higgs VEVs must also be real.

An additional constraint originates from two minimization conditions, $\frac{\partial V}{\partial v_R}=\frac{\partial V}{\partial\mathfrak{Re}v_L}=0$, and is known as the 'VEV seesaw relation':
\begin{align}
\beta_2k_1^2+\beta_1k_1k_2+\beta_3k_2^2=(2\rho_1-\rho_3)v_Lv_R ~,
\label{eq.VEVseesaw}
\end{align}
which gives
\begin{align}
v_L=\gamma\,\frac{k_1^2+k_2^2}{v_R}~,
\end{align}
where
\begin{align}
\gamma \equiv \frac{\beta_2k_1^2+\beta_1k_1k_2+\beta_3k_2^2}{(2\rho_1-\rho_3)(k_1^2+k_2^2)}~.
\end{align}
Assuming that  $\beta_i$ and $\rho_i$ are of order unity (i.e. not too large - to preserve
unitarity, and not too small - to avoid fine tuning) implies that $\gamma\sim 1$.
Since the light neutrino masses (which are proportional to $v_L$ via the Yukawa coupling)
are bound to be less
than ${\cal O}(1)\;\unit[]{eV}$ \cite{neutrinomass}, $v_R$ has to be at least as large as ${\cal O}(10^8)$ GeV.
This, in turn, leads to unobservably large masses
for the additional Higgs and gauge bosons states (i.e., of order $10^8$ GeV),
unless the $\beta_i$ are fine-tuned to
reduce $\gamma$ to about $10^{-7}$ for which $v_R$ can be considerably smaller, i.e., $v_R \sim {\cal O}(10^3)$ GeV.
In this case, the new gauge-bosons and Higgs particles become accessible at the LHC
(see mass formulae in \ref{App:AppendixB}). One possible way to avoid the (unwanted) fine-tuning of the Higgs couplings is to eliminate almost completely the VEV seesaw relation by setting some of the relevant parameters in the Higgs potential, and in particular the $\beta$ parameters, to zero. This may not be considered as fine tuning but, rather, as a possible consequence of some yet higher symmetry (e.g., GUT or SUSY), which lies beyond the context of the LRSM \cite{classic}.\footnote{Setting the $\beta_i$ parameters to be small but not zero through mechanisms such as horizontal symmetry is discussed e.g., in \cite{kiers}. However here we follow the stricter constraint presented in Refs.\cite{classic,gluza}.}
Following this approach, the VEV see-saw relation reduces to a ``remnant" VEV see-saw relation
\begin{align}
(2\rho_1-\rho_3)\,v_L\,v_R=0 ~.
\label{eq.VEVseesaw-rem}
\end{align}
The only way for Eq.\eqref{eq.VEVseesaw-rem} to be consistent with observation is to set the VEV of the left-handed triplet, $v_L$, to zero \cite{classic2}.\footnote{The other option of setting $v_R=0$ leads to
$m_{W_2} \sim m_W$ in contrast with observation,
and the option $2\rho_1-\rho_3=0$ leads to massless Higgs bosons of the left-handed triplet,
which is also ruled out experimentally\cite{conference}.}

Summarizing the above, the following constraints are imposed due to requiring explicit CP-conservation, requiring natural grounds to support the minimization conditions, and requiring consistency with experiment:
\begin{itemize}
\item
All the Higgs multiplets VEVs are real.
\item
The parameters $\beta_i$ are set to zero.
\item
The Higgs left triplet VEV $v_L$ is set to zero.
\end{itemize}

Moving on to the Higgs mass content, its mass matrix is determined by
\begin{align}
\frac{\partial^2}{\partial\phi_i\partial\phi_j}\,V\,\Big|_{\phi_i=\phi_j=0} = m^2_{i,j}.
\label{eq.mass-matrix}
\end{align}
Imposing the above three constraints simplifies the diagonalization procedure of the Higgs mass matrix, whose precise form is given in \cite{classic}. The expressions for the Higgs eigenmasses in terms of the parameters in the Higgs potential and the Higgs VEVs are given in \ref{App:AppendixB} for the case $v_R \gg k_{1,2}$, as required phenomenologically.\footnote{As mentioned above, heavy gauge bosons haven't been discovered yet, and thus should be significantly heavier than the SM $W$, which mass is proportional to $\sqrt{k_1^2+k_2^2}$. For lower mass limits of $W_R$ see footnote \ref{note0}.} The Higgs fields which comprise the multiplets (referred to as the non-physical Higgs fields) are presented in terms of the Higgs eigenstates in \ref{appendix.Higgs physical eigenstates}.

\subsection{Obtaining the Yukawa terms in the physical basis }
\label{sub.yukawa}
The Yukawa terms given in Eq.\eqref{eq.yukawa} are given in terms of gauge eigenstates and Yukawa matrices, whereas a more suitable form for calculations depends on physical basis of mass eigenstates and empirical parameters as explained in sec.\ref{subsec.Lag-struc}. We describe here the structure of the Higgs-quark and the Higgs-lepton sectors and their conversion into this preferred basis. We will describe the implementation of the Yukawa sector in FeynRules in the next section.

When considering the most general way in which the above Higgs multiplets can be coupled to bilinear fermion field products to form singlets under $SU(2)_L\times SU(2)_R \times U(1)$, one should also bear in mind the Majorana type lepton-Higgs couplings which complement the Dirac type ones. We write again, for clarity, the general Yukawa terms of Eq.\eqref{eq.yukawa}:
\begin{align}
L_Y=-\Sigma_{i,j}\big[&\bar{L}_{iL}((h_L)_{ij}\phi+(\tilde{h_L})_{ij}\tilde{\phi})L_{jR} - \bar{Q}_{iL}((h_Q)_{ij}\phi+(\tilde{h_Q})_{ij}\tilde{\phi})Q_{jR} \nonumber \\
& -\overline{\strut{(L_{iR})}^c}\;\Sigma_R {(h_M)}_{ij} L_{jR} -\overline{\strut{(L_{iL})}^c}\;\Sigma_L {(h_M)}_{ij}L_{jL} \big]
+h.c.
\label{eq.yukawa2}
\end{align}
where, as mentioned above, $\tilde{\phi}\,\equiv\, \tau_2\phi^*\tau_2$, $\Sigma_{L,R}=i\tau_2\Delta_{L,R}$ and $h_Q$, $h_L$,$h_M$,$\tilde{h_Q}$, $\tilde{h_L}$ are $3\times 3$ Yukawa matrices in flavor space ($h_M$ is symmetric). These matrices can be defined in terms of VEVs, diagonal mass matrices and fermion mixing matrices, thus yielding the desired Feynman rules. We shall now demonstrate the process.

Starting with the quark sector, the Yukawa terms for the coupling with $\phi$-type Higgs fields are
\begin{align}
& -\bar{U'}_L (h_Q\phi^0_1+\tilde{h}_Q\phi^{0*}_2) U'_R\;-\;\bar{D'}_L(h_Q\phi^0_2+\tilde{h}_Q\phi^{0*}_1)D'_R \nonumber \\
& -\bar{U'}_L (h_Q\phi^+_1-\tilde{h}_Q\phi^+_2) D'_R\;-\;\bar{D'}_L(h_Q\phi^-_2+\tilde{h}_Q\phi^-_1)U'_R\;+\;\text{h.c.}
\label{eq.yukawa-quark}
\end{align}
where $U'_{L,R},D'_{L,R}$ are three dimensional vectors built of up-type and down-type quark gauge eigenstates, respectively (e.g $\bar{U}_L=(\bar{u}'_L\bar{c}'_L\bar{t}'_L)$). After SSB the mass terms which arise are
\begin {equation}
-\bar{U}'_LM_uU'_R-\bar{D}'_LM_dD'_R+h.c., \label{masslagrangian}
\end{equation}
where
\begin{align}
M_u\equiv\frac{1}{\sqrt{2}}\left(h_Qk_1+\tilde{h}_Qk_2\right),\hspace{0.2cm} M_d \equiv \frac{1}{\sqrt{2}}\left(\tilde{h}_Qk_1+h_Qk_2\right).
\end{align}

Because the VEVs $k_1$ and $k_2$ are real in the considered model (since, as discussed above, this model does not contain explicit CP violation) and, in addition, $h_Q$ and $\tilde{h_Q}$ are hermitian (as implied from the left right symmetry), it implies that $M_u$ and $M_d$ are also hermitian. These matrices can therefore be diagonalized by unitary transformations to give up-type and down-type, real, diagonal mass matrices
\begin{align}
M^u_{diag} \equiv V^{u\dagger}_L M_u V^u_R=diag(m_u,m_c,m_t), \hspace{5 mm} M^d_{diag} \equiv V^{d\dagger}_L M_d V^d_R=diag(m_d,m_s,m_b),
\end{align}
where the unitary matrices $V^u_{L,R}$ and $V^d_{L,R}$ rotate the gauge eigenstates into mass eigenstates,
\begin{equation}
U'_{L,R}=V^u_{L,R}U_{L,R},\hspace{5 mm} D'_{L,R}=V^d_{L,R}D_{L,R}.
\label{eq.unitary-quarks}
\end{equation}
The diagonal (and real) mass matrices can then be used to express the Yukawa matrices
\begin{align}
& h_Q = \frac{\sqrt{2}}{k_-^2}\left(k_1 V_L^u M^u_{diag} V_R^{u\dagger}-k_2V_L^d M^d_{diag} V_R^{d\dagger}\right), \nonumber \\
& \tilde{h}_Q=\frac{\sqrt{2}}{k_-^2}\left(-k_2 V_L^u M^u_{diag} V_R^{u\dagger}+k_1V_L^d M^d_{diag} V_R^{d\dagger}\right).
\label{eq.yukawa-matrices}
\end{align}
where $k_-^2= k_1^2-k_2^2$. Upon inserting this result into the term for up-type quark coupling with neutral $\phi$-type Higgs fields for example, which is given by (see Eqs.\eqref{eq.yukawa-quark} and \eqref{eq.unitary-quarks})
\begin{align}
\bar{U}_LV^{u\dagger}_L(h_Q\phi^0_1+\tilde{h}_Q\phi^{0*}_2)V^u_RU_R\;+\;\text{h.c.}
\end{align}
one obtains the Yukawa interaction in the up-quark sector
\begin{align}
-\frac{\sqrt{2}}{k_-^2}\bar{U}_L\,\left(M^u_{diag}(k_1\phi^0_1-k_2\phi^{0*}_2)+U^{CKM}_L M^d_{diag} U^{CKM\dagger}_R(-k_2\phi^0_1+k_1\phi^{0*}_2)\right)\,U_R\;+\;\text{h.c.},
\label{eq.Yukawa-quark-final}
\end{align}
where the left and right CKM matrices, $U^{CKM}_L$ and $U^{CKM}_R$, are given by
\begin{equation}
U^{CKM}_L\equiv V^{u\dagger}_LV^d_L,\hspace{3 mm}U^{CKM}_R\equiv V^{u\dagger}_RV^d_R.
\label{eq.CKM}
\end{equation}

Expressing the non-physical Higgs states in Eq.\eqref{eq.Yukawa-quark-final} in terms of physical states (see \ref{appendix.Higgs physical eigenstates}) yields the general interaction terms in a form which depends on physical states and parameters.

At this point, a brief explanation about the implication of the manifest / quasi-manifest nature of the left-right model on the relative definition of the left and right CKM matrices is in place. In the absence of explicit CP violation in the Higgs potential, the diagonalization of the up and/or down mass matrices may give negative diagonal mass terms. This is known as quasi-manifest left-right model (as opposed to the manifest model where the diagonalization results solely in positive masses). In this case it is useful to define
\begin{align}
& V^u_R = V^u_L W^u, \nonumber \\
& V^d_R = V^d_L W^d
\label{eq.m-qm}
\end{align}
where $W_u$ and $W_d$ are diagonal $3\times 3$ matrices with $\pm 1$ as diagonal elements, aimed to set the sign of each element in the up type and down type diagonal mass matrix, respectively, to be positive. The relation between the two CKM matrices can thus be written as
\begin{align}
U^{CKM}_R=W^U \, U^{CKM}_L \, W^D.
\label{eq.ckm}
\end{align}

Turning now to the lepton sector, the Yukawa terms for the neutrino and the charged lepton couplings with the Higgs fields are
\begin{align}
& -\bar{\nu'}_L (h_L\phi^0_1+\tilde{h}_L\phi^{0*}_2) \nu'_R-\bar{l'}_L(h_L\phi^0_2+\tilde{h}_L\phi^{0*}_1)l'_R-\bar{\nu'}_L (h_L\phi^+_1-\tilde{h}_L\phi^+_2) l'_R \nonumber \\
& -\bar{l'}_L(h_L\phi^-_2+\tilde{h}_L\phi^-_1)\nu'_R + \big\{-\overline{(\nu'^c)}_R h_M \delta_L^0 \nu'_L+\overline{(l'^c)}_R h_M \delta^{++}_L l'_L +\frac{1}{\sqrt{2}}\overline{(\nu'^c)}_R h_M \delta_L^+ l'_L \nonumber \\
&  +\frac{1}{\sqrt{2}} \overline{(l'^c)}_R h_M \delta_L^+\nu'_L+(L\leftrightarrow R) \big\} + \text{h.c.}
\label{eq.yukawa-lepton}
\end{align}
where $\nu'_{L,R}$ and $l'_{L,R}$ are three dimensional vectors built of the neutrino and charged lepton gauge eigenstates, respectively. After SSB, the neutrino and charged lepton mass terms which stem from the above Yukawa terms are
\begin{align}
&L^{lepton}_{mass} =-\frac{1}{2}(\overline{n'_L} M_\nu n'_R+\bar{n}'_R M^{*}_\nu n'_L)-(\bar{l}'_LM_ll'_R+\bar{l}'_{R}M^{\dagger}_ll'_L),
\label{eq.lepton-mass}
\end{align}
where
\begin{align}
n'_R=\left(\begin{array}{c}(\nu'^{c})_R\\ \nu^{'}_R\end{array} \right),n'_L=\left(\begin{array}{c}\nu'_L\\ (\nu^{'c})_L\end{array}\right),
\label{eq.yukawa-lepton2}
\end{align}
and recalling that 
\begin{align}
(\psi^c)_{L,R}=(\psi_{R,L})^c
\label{eq.identity}
\end{align}
where $\psi$ is a fermion field.

The Yukawa mass matrix of the charged leptons results from the VEVs of the bidoublet. As with the quark mass matrices, it consists of Dirac-type mass terms
\begin{align}
M_l=\frac{1}{\sqrt{2}}\left(h_L k_2+\tilde{h}_L k_1\right)=M^{\dagger}_l
\label{eq.charged-lepton-mass}
\end{align}
(it is also hermitian, as explained in the case of quark mass matrices). This matrix can be diagonalized by a unitary transformation
\begin{align}
l'_{L,R}=V^l_{L,R}l_{L,R},
\label{eq.lepton-states}
\end{align}
which enables one to define a diagonal, real, charged lepton mass matrix
\begin{equation}
M^l_{diag} \equiv V^{l\dagger}_L M_l V^l_R=diag(m_e,m_\mu,m_\tau).
\label{eq.charged-lepton-mass-diagonal}
\end{equation}
The $6 \times 6$ Yukawa mass matrix of the neutrinos consists of both Majorana type and Dirac type couplings:
\begin{align}
&M_\nu=\left(
\begin{array}{cc}
M_L&M_D\\M_D^T&M_R\end{array}\right)
\label{eq.massmatrix}
\end{align}
where the mass matrix $M_\nu$ is symmetric, $M_D$ is a Dirac-type block, and $M_L$ and $M_R$ are Majorana-type blocks). The diagonalization of the neutrino Yukawa mass matrix is done by a single $6 \times 6$ unitary transformation $V$:
\begin{align}
M^{\nu}_{diag} \equiv V^TM_{\nu}V ~,
\label{eq.diag1}
\end{align}
Substituting eq.\eqref{eq.diag1} in eq.\eqref{eq.lepton-mass} gives
\begin{align}
n'_L=V^*N_L,\;n'_R=VN_R ~,
\label{eq.neutrino-eigenstates}
\end{align}
where $N_{R,L}$ are the right and left handed projections, respectively, of the six mass eigenstates of the Majorana neutrinos. The
matrix $V$ can be conveniently defined as
\begin{equation}
V=\left(\begin{array}{c}V^{\nu*}_L\\V^{\nu}_R\end{array}\right),
\label{eq.conveniently-defined}
\end{equation}
which implies the following neutrino mixings:
\begin{align}
\nu'_R=V^\nu_R N_R,\quad \nu'_L=V^\nu_L N_L.
\label{eq.neutrino-unitary}
\end{align}
Some useful identities that will be used below can be derived from the unitarity of V
\begin{align}
V V^\dagger=\left(\begin{array}{c}V^{\nu *}_L \\ V^\nu_R \end{array} \right) \left(\begin{array}{cc} V^{\nu T}_L & V^{\nu\dagger}_R \end{array} \right) =\left(\begin{array}{cc} V^{\nu *}_L V^{\nu T}_L & V^{\nu *}_L V^{\nu\dagger}_R \\ V^\nu_R V^{\nu T}_L & V^\nu_R V^{\nu\dagger}_R \end{array} \right)=\mathbb{1}
\end{align}
which implies
\begin{align}
V^\nu_LV^{\nu \dagger}_L=\mathbb{1}, \quad V^\nu_R V^{\nu T}_L=0 \quad V^\nu_R V^{\nu\dagger}_R=\mathbb{1}.
\label{eq.useful1}
\end{align}
Using the above formalism (in particular eqs.\eqref{eq.neutrino-eigenstates} and \eqref{eq.conveniently-defined}) and substituting the neutrino mass matrix of eq.\eqref{eq.massmatrix} into eq.\eqref{eq.lepton-mass}, it is possible to rewrite for example
$-\bar{n}'_LM_\nu n'_R$ as
\begin{align}
-\bar{n}'_LM_\nu n'_R=-\bar{N}_L\underbrace{V^TM_\nu V}_{M^\nu_{\text{diag}}} N_R = & -\bar{N}_L V^{\nu \dagger}_LM_D V^\nu_R N_R-\bar{N}_L V^{\nu T}_RM_D^T V^{\nu*}_L N_R \nonumber \\
&-\bar{N}_L V^{\nu T}_R M_R V^\nu_R N_R -\bar{N}_L V^{\nu \dagger}_L M_L V^{\nu *}_L  N_R,
\end{align}
and thus obtain
\begin{align}
M^\nu_{\text{diag}}=V^{\nu \dagger}_L M_D V^\nu_R+V^{\nu T}_R M_D^T V^{\nu *}_L+V^{\nu T}_R M_R V^\nu_R+V^{\nu \dagger}_L M_L V^{\nu *}_L.
\end{align}
Using in addition the identities of Eq.\eqref{eq.useful1} one gets
\begin{align}
& M_D=V^\nu_LM^\nu_{\text{diag}}V^{\nu\dagger}_R=M_D^T, \nonumber \\
& M_R=V^{\nu *}_RM^\nu_{\text{diag}}V^{\nu\dagger}_R=M_R^T, \nonumber \\
& M_L=V^\nu_LM^\nu_{\text{diag}}V^{\nu T}_L=M_L^T.
\label{eq.useful2}
\end{align}
These mass matrix blocks may also be written in terms of Yukawa matrices and Higgs VEVs.
In particular, the Dirac mass block $M_D$ originates from the bidoublet VEVs
\begin{align}
M_D=\frac{1}{\sqrt{2}}\left(h_Lk_1+\tilde{h}_Lk_2\right)
\label{eq.dirac-matrix}
\end{align}
and is real (in light of Eq.\eqref{eq.useful2}). The Majorana mass block $M_L$ arises from the VEV of the left handed Higgs triplet:
\begin{align}
M_L=\sqrt{2}h_M v_L ~,
\label{eq.left-matrix}
\end{align}
and the Majorana mass block $M_R$ is generated by the right handed Higgs triplet VEV:
\begin{equation}
M_R=\sqrt{2}h_Mv_R.\footnote{Since the implemented model uses real neutrino mass eigenvalues, $h_M$ can be regarded as real. This is consistent with the assumption made in \cite{classic}.}
\label{eq.right-matrix}
\end{equation}

Before transforming the lepton Yukawa matrices to physical basis, it is of importance to relate the above discussed VEV see-saw relation to the neutrino mass see-saw relation. This can be done by focusing on a single generation neutrino mass system (for simplicity, we ignore generation mixing). The Diagonalization of this system can be conveniently done by defining the self conjugate (Majorana) spinors
\begin{align}
f\equiv\nu'_L+(\nu'_L)^c,\;\;\; F\equiv\nu'_R+(\nu'_R)^c
\end{align}
where $\psi^c\equiv C\bar{\psi}^T$.
Using this, the neutrino Yukawa mass terms of Eq.\eqref{eq.lepton-mass} can be written as
\begin{align}
\frac{1}{2}\left( \begin{array}{c} \bar{f} \\ \bar{F} \end{array} \right)^T \left( \begin{array}{cc} M_L & M_D \\ M_D & M_R \end{array} \right)
\left( \begin{array}{c} f \\ F \end{array} \right).
\end{align}
As mentioned earlier, the ``remnant" VEV see-saw relation of Eq.\eqref{eq.VEVseesaw-rem} implies the constraint
$v_L=0$, so that, in this limit and with the approximation $v_R \gg k_1,k_2$, the neutrino mass eigenvalues are (see Eqs.\eqref{eq.dirac-matrix}-\eqref{eq.right-matrix})
\begin{align}
m_{\nu}\simeq-\frac{M_D^2}{M_R},\;\;\; m_{N}\simeq M_R,
\end{align}
where $m_{\nu}$ and $m_{N}$ are the light and the heavy Majorana neutrino masses, respectively. The corresponding approximate Majorana eigenstates are
\begin{align}
& \nu\simeq f-\frac{M_D}{M_R}F, \\
& N\simeq F+\frac{M_D}{M_R}f ~,
\label{eq.majorana-eigenstates}
\end{align}
and the product of the two mass eigenvalues is
\begin{align}
m_{\nu} m_{N} = -M_D^2 ~.
\end{align}
This is the widely known (type I) see-saw relation. A special case can be reached by assuming that the neutrino Dirac mass $M_D$ is of the same order as the related charged lepton (this occurs e.g. when $\tilde{h}_L \sim h_L$ and $k_1 \sim k_2$ \cite{classic}), and thus one gets
\begin{align}
m_{\nu} m_{N} = -m_l^2,
\label{eq.specialseesaw}
\end{align}
where $m_l$ is the mass of the charged lepton.\footnote{Redefining the $\nu$ eigenstate as $\nu \rightarrow \gamma^5(f-\frac{M_D}{M_R}F)$ changes the sign of the corresponding mass eigenvalue (see \cite{kayser}) in which case the sign in the right hand side of Eq.\eqref{eq.specialseesaw} becomes positive.}

Expressing the charged current terms in Eq.\eqref{lag.fermions} in terms of the above Majorana fields gives
\begin{align}
& \bar{e}'_L\gamma_\mu\nu'_L=\bar{e}'_L\gamma_\mu f_L\simeq \bar{e}'_L\gamma_\mu\left[\nu+\frac{M_D}{M_R}N\right]_L, \nonumber \\
& \bar{e}'_R\gamma_\mu\nu'_R=\bar{e}'_R\gamma_\mu F_R\simeq \bar{e}'_R\gamma_\mu\left[N-\frac{M_D}{M_R}\nu\right]_R.
\label{charged-current-majorana}
\end{align}
The coefficients of the eigenstates $\nu$ and $N$ in this equation are CKM-type lepton mixings, which will appear again as matrices in the three generation case below.\footnote{We ignored, in this single generation example, the charged lepton inner mixings (the $V^l_{L,R}$ matrix of Eq.\eqref{eq.lepton-states}), which are part of the general discussion above (and should of course be included in the CKM-type mixings). The model implementation file, however, allows the user to include the charged lepton mixings as well, as explained in the implementation section.}

 Going back to obtaining a physical basis for the Yukawa terms, it is now possible, using the expressions for $M_D$ (Eqs.\eqref{eq.useful2} and \eqref{eq.dirac-matrix}) and for the charged lepton mass matrix (Eqs.\eqref{eq.charged-lepton-mass} and \eqref{eq.charged-lepton-mass-diagonal}), to extract the values of the lepton Yukawa matrices $h_L$ and $\tilde{h}_L$ in terms of the lepton mass values, the Higgs VEVs and the mixings :
\begin{align}
& h_L=\frac{\sqrt{2}}{k_-^2}\left(k_1 V_L^\nu M^\nu_{\text{diag}} V_R^{\nu\dagger}-k_2V^l_L M^l_{\text{diag}} V^{l\dagger}_R\right), \nonumber \\
& \tilde{h}_L=\frac{\sqrt{2}}{k_-^2}\left(-k_2V_L^\nu M^\nu_{\text{diag}} V_R^{\nu\dagger}+k_1 V^l_L M^l_{\text{diag}} V^{l\dagger}_R\right).
\label{eq.hL}
\end{align}
Furthermore, using the expressions for $M_R$ (Eqs.\eqref{eq.useful2} and Eqs.\eqref{eq.right-matrix}) one can obtain the Yukawa matrix $h_M$:
\begin{align}
h_M=\frac{1}{\sqrt{2}\nu_R}\,V^{\nu *}_RM^\nu_{\text{diag}}V^{\nu\dagger}_R.
\label{eq.hm}
\end{align}
Inserting the above expressions into the leptonic sector of the Yukawa Lagrangian is done in the same way as in the quark sector. Focusing for example on the left handed singly charged Higgs terms,
\begin{align}
& \frac{1}{\sqrt{2}}\overline{(\nu'^c)}_R h_M \delta^+_L l'_L  + \frac{1}{\sqrt{2}} \overline{(l'^c)}_R h_M \delta_L^+ \nu'_L\text{ + h.c.} \nonumber \\*
=&\frac{1}{2\nu_R}\big[\bar{N}_RV^{\nu T}_LV^{\nu*}_RM^\nu_{\text{diag}}V^{\nu\dagger}_RV^l_Ll_L
\delta^+_L+\overline{(l^c)_R}V^{lT}_L V^{\nu*}_RM^\nu_{\text{diag}}V^{\nu\dagger}_R V^\nu_L N_L
\delta^+_L\big]\text{ + h.c.} \nonumber \\*
=&\frac{1}{\nu_R}\big[\bar{N}_RV^{\nu T}_LV^{\nu*}_RM^\nu_{\text{diag}}V^{\nu\dagger}_RV^l_Ll_L
\delta^+_L\big]\text{ + h.c.}. 
\label{eq.Yukawa-lepton-final}
\end{align}
where Eqs.\eqref{eq.yukawa-lepton2}, \eqref{eq.identity}, \eqref{eq.lepton-states} and \eqref{eq.neutrino-unitary} were used, as was also the identity
\begin{align}
\overline{(l^c)}_R N_L=\bar{N}_R l_L
\end{align}
(the creation phase factor for the Majorana neutrino field is chosen to be 1, i.e. $N=N^c$\cite{gluza}\footnote{The Majorana field is $\Psi^M(x)\propto \sum_{\textbf{p},s} \big[f(\textbf{p},s) u(\textbf{p},s)e^{ipx}+\lambda f^\dagger(\textbf{p},s)v(\textbf{p},s)e^{-ipx}\big]$ (see \cite{kayser}). The creation phase factor, $\lambda$, is chosen in our work and in \cite{gluza} to be 1.}). The final result of Eq.\eqref{eq.Yukawa-lepton-final} can be written in terms of CKM-type mixing matrices in the leptonic sector,
\begin{align}
K_L=V^{\nu\dagger}_L V^l_L ~, \nonumber \\
K_R=V^{\nu\dagger}_R V^l_R ~,
\label{eq.kl-kr}
\end{align}
which are $6\times3$ matrices - analogues to the CKM mixing matrices in the quark sector. The lepton sector comprises a quasi-manifest control matrix as well, denoted as $W^l$
\begin{align}
V^l_L=V^l_R W^l,
\label{eq.quasi-lepton}
\end{align}
which is the analogue of the $W^d$ matrix in the quark sector.

Using the above definitions, Eq.\eqref{eq.Yukawa-lepton-final} can be rewritten as
\begin{align}
\frac{1}{\nu_R}\big[\bar{N}_RK^*_LW_lK^T_RM^\nu_{\text{diag}}K_RW_ll_L
\delta^+_L\big]+\text{h.c.}
\label{eq.Yukawa-lepton-final2}
\end{align}
so that, similarly to the Higgs-quark interactions,
expressing the non-physical Higgs states of Eq.\eqref{eq.Yukawa-lepton-final2} in terms of physical states (see \ref{appendix.Higgs physical eigenstates}) gives the general Higgs-lepton interactions in the physical basis.

\subsection{The gauge boson eigensystem}
\label{sub.eigensystem}

As usual, the interactions of the Higgs multiplets with the gauge bosons in $\mathcal{L}_{Higgs}$ (see Eq.\eqref{eq.scalarL}) arise from the covariant derivatives. After SSB, the terms which contain a bilinear product of two gauge boson fields give rise to two mass matrices, a $2\times2$ matrix and $3\times 3$ matrix, which correspond to the charged $W_L^\pm$, $W_R^\pm$ ($W_{L,R}^{\pm\mu}=\frac{1}{\sqrt{2}}(W_{L,R}^{1\mu}\mp i W_{L,R}^{2\mu})$) gauge boson system and the neutral $(W_L^3$, $W_R^3,B)$ gauge boson system, respectively. Thus the mass terms are given by \cite{gluza}
\begin{equation}
L_M=\left(W_L^{+\mu},W_R^{+\mu}\right)\tilde{M}_W^2\left(\begin{array}{c}
W^{-}_{L\mu}\\
W^{-}_{R\mu} \end{array}\right)+h.c.+\frac{1}{2}\left(W^{\mu}_{3L},W^{\mu}_{3R},B^{\mu}\right)\tilde{M}_0^2\left(\begin{array}{c}
W_{3L\mu}\\
W_{3R\mu}\\
B_{\mu} \end{array}\right), \label{gaugemass}
\end{equation}
where the mass matrices are (assuming $v_L$=0, see the above discussion)
\begin {equation}
\tilde{M}_W^2=\frac{g^2}{4}\left(\begin{array}{cc}
k_{+}^2&-2k_1k_2\\
-2k_1k_2&k^2_{+}+2v_R^2 \end{array} \right),
\end{equation}
and
\begin {equation}
\tilde{M}_0^2=\frac{1}{2}\left(\begin{array}{ccc}
\frac{g^2}{2}k_{+}^2&-\frac{g^2}{2}k_{+}^2&0\\
\frac{-g^2}{2}k_{+}^2&-\frac{g^2}{2}(k_{+}^2+v_R^2)&-2gg'v^2_R\\
0&-2gg'v_R^{2}&2g'^2v^2_R \end{array} \right),
\end{equation}
where 
\begin{align}
&g=g_L=g_R=\frac{e}{\sin{\Theta_W}},\quad g'=\frac{e}{\sqrt{{\cos{2\Theta_W}}}},
\end{align}
and $k_+ \equiv \sqrt{k_1^2+k_2^2}$ ($k_1\text{ and }k_2$ are real).\\
The symmetric mass matrices are diagonalized by the orthogonal transformations
\begin{equation}
\left(\begin{array}{c}
W_L^{\pm}\\
W_R^{\pm} \end{array} \right)=\left(\begin{array}{cc}
\cos\xi&\sin\xi\\
-\sin\xi&\cos\xi \end{array} \right)=\left(\begin{array}{c}
W_1^{\pm}\\
W_2^{\pm} \end{array} \right),
\label{chargedbosons}
\end{equation}
and
\begin{eqnarray}
&\left(\begin{array}{c}
W_{3L}\\W_{3R}\\B\end{array}\right)=
\left(\begin{array}{ccc} c_W c & c_W s & s_W \\
-s_W s_M c-c_M s & -s_Ws_Ms+c_Mc&c_Ws_M\\-s_Wc_Mc+s_Ms&-s_Wc_Ms-s_Mc&c_Wc_M \end{array}\right)\left(\begin{array}{c}Z_1\\Z_2\\A\end{array}\right)
\label{neutralbosons}
\end{eqnarray}
where
\begin{align}
&c_W=cos{\Theta_W}, s_W=sin{\Theta_W}, c_M=\frac{\sqrt{\cos{2\Theta_W}}}{\cos{\Theta_W}}, \nonumber \\
& s_M=\tan{\Theta_W}, c=\cos{\phi}, s=\sin{\phi} ~, \nonumber
\end{align}
and the explicit expressions for the charged and neutral mixing angles $\xi$ and $\phi$ are shown in \ref{App:AppendixB}. The masses of the physical gauge bosons are given by
\newpage
\begin{align}
& M_{W_{1,2}}^2=\frac{g^2}{4}\big[k_+^2+v_R^2 \mp \sqrt{v_R^4+4\,k_1^2\,k_2^2}\,\big],\nonumber \\
& M_{Z_{1,2}}^2=\frac{1}{4}\big[\big[g^2 k_+^2+2v_R^2(g^2+g'^2)\big] \mp\,\sqrt{\big[g^2k_+^2+2v_R^2(g^2+g'^2)\big] ^2-4g^2(g^2+2g'^2)\,k_+^2v_R^2}\big].
\label{eq.W-bosons}
\end{align}
\section{The model implementation}
\label{sec.implement}
\subsection{Fermion fields, mixing matrices, Yukawa terms}
The model file defines doublets of fermion gauge eigenstates using CKM type mixing matrices and ascribes the mixing of the mass eigenstates completely to either $T_3=-\frac{1}{2}$ or $T_3=+\frac{1}{2}$ isospin states. This can be seen for example in the definition of the left-handed quark doublet:
\begin{verbatim}
QL[sp1_,1,ff_,cc_]:>	Module[{sp2}, ProjM[sp1,sp2] uq[sp2,ff,cc]],
QL[sp1_,2,ff_,cc_]:>	Module[{sp2,ff2},CKML[ff,ff2]
                    ProjM[sp1,sp2] dq[sp2,ff2,cc]],
\end{verbatim}
where  \verb+ProjM+ denotes the left handed projection operator, \verb+sp1+ and \verb+sp2+ are spinor indices, \verb+ff+ and \verb+ff2+ are generation indices and \verb+cc+ is a color index. The above definition keeps the $+\frac{1}{2}$ isospin states as unmixed left handed projections of the up type quarks mass eigenstates \verb+uq+, and assigns the product of the up and down mixing matrices, i.e. the left handed CKM matrix (defined in Eq.\eqref{eq.CKM} and denoted as \verb+CKML+) to the left handed projections of the down type quarks mass eigenstates (denoted as \verb+dq+), i.e. to the $-\frac{1}{2}$ isospin states. The lepton doublets are implemented in a similar way, with the only difference being in assigning the mixing to the $+\frac{1}{2}$ isospin states (the neutrino fields) by using the CKM type $K_L$ and $K_R$ leptonic mixing matrices as defined in Eq.\eqref{eq.kl-kr}. For example,
the left handed lepton doublet is given by
\begin{verbatim}
LL[sp1_,1,ff_] :> Module[{sp2,ff2}, Conjugate[KL[ff2,ff]]
                  ProjM[sp1,sp2] Nl[sp2,ff2]],
LL[sp1_,2,ff_] :> Module[{sp2}, ProjM[sp1,sp2] l[sp2,ff]]
\end{verbatim}
where, in addition to the symbols used in the \verb+QL+ definition above, \verb+Nl+ denotes the neutrino mass eigenstates, \verb+l+ denotes the charged lepton mass eigenstates and \verb+KL+ denotes the left handed lepton mixing matrix (defined in Eq.\eqref{eq.kl-kr}). Furthermore, charge conjugate lepton doublet projections were defined for the lepton-$\Delta_{R,L}$ interactions. For example, the left handed projection of the lepton doublet charge conjugate is defined as
\begin{verbatim}
LCL[sp1_,1,ff_] :> Module[{sp2,ff2}, KR[ff2,ff]
                   ProjM[sp1,sp2] Nl[sp2,ff2]],
LCL[sp1_,2,ff_] :> Module[{sp2}, ProjM[sp1,sp2] CC[l[sp2,ff]]]
\end{verbatim}
the notation \verb+LCL+ denotes $(L_i^c)_L$ which (using Eq.\eqref{eq.identity}) is given by
\begin{align}
(L_i^c)_L & =\left( \begin{array}{c} \nu'^c_i \\
    l^c_i \end{array} \right)_L=
 \left( \begin{array}{c} (\nu'_{iR})^c \\
    (l_i^c)_L \end{array} \right) =
\left( \begin{array}{c} (K^\dagger_{Rij} N_{jR})^c \\[0.1cm]
    (l^c_i)_L \end{array} \right) = \left( \begin{array}{c} K^T_{Rij} N^c_j \\[0.1cm]
    l^c_i \end{array} \right)_L \nonumber \\[5pt]
    & = \left( \begin{array}{c} K^T_{Rij} N_j \\[0.1cm]
    l^c_i \end{array} \right)_L
\end{align}
where $i$ is the lepton generation index, and $j\;(=1..6)$ is a Majorana neutrino index.

Using this formalism when implementing the Yukawa interactions requires some caution however, as we shall now demonstrate. Starting with the quark couplings, the expressions for the Yukawa quark matrices $h_Q$ and $\tilde{h}_Q$ given in Eq.\eqref{eq.yukawa-matrices} transform, when deriving them using the above formalism, into
\begin{align}
& h_Q = \frac{\sqrt{2}}{k_-^2}\left(k_1 M^u_{diag}-k_2 U^{CKM}_L M^d_{diag} U^{CKM\dagger}_R \right), \nonumber \\
& \tilde{h}_Q=\frac{\sqrt{2}}{k_-^2}\left(-k_2 M^u_{diag} +k_1 U^{CKM}_L M^d_{diag} U^{CKM\dagger}_R\right).
\end{align}
This is implemented as
\begin{Verbatim}[samepage=true]
 yQ[a_,b_] :> Module[{sp5,sp6}, Sqrt[2]/(k1^2-k2^2) (k1 yMU[a,b]
              - k2 CKML[a,sp5] yDO[sp5,sp6] HC[CKMR[b,sp6]])]
 yQtilde[a_,b_] :> Module[{sp5,sp6}, Sqrt[2]/(k1^2-k2^2)(-k2 yMU[a,b]
                   + k1 CKML[a,sp5] yDO[sp5,sp6] HC[CKMR[b,sp6]])]
\end{Verbatim}
where \verb+yQ+ and \verb+yQtilde+ denote the Yukawa matrices $h_Q$ and $\tilde{h}_Q$, respectively, \verb+yMU+ and \verb+yMD+ are the diagonal up-quark and down-quark matrices, respectively, \verb+k1+ and \verb+k2+ are the Higgs VEVs, and \verb+CKML+ and \verb+CKMR+ are the left handed and right handed mixing CKM matrices, respectively. Implementing the quark doublets and the quark Yukawa matrices in above formalism and substituting them in the quark-Higgs Yukawa term (see Eq.\eqref{eq.yukawa}) reproduces the same Feynman rules derived in sec.\ref{sub.yukawa}.

A similar procedure is done with the $\phi$-type lepton-Higgs interactions, where the $h_L$ and $\tilde{h}_L$ Yukawa matrices (given in Eq.\eqref{eq.hL}) become
\begin{align}
& h_L=\frac{\sqrt{2}}{k_-^2}\left(k_1 K^\dagger_L M^\nu_{\text{diag}} K_R-k_2 M^l_{\text{diag}} \right), \nonumber \\
& \tilde{h}_L=\frac{\sqrt{2}}{k_-^2}\left(-k_2 K_L^\dagger M^\nu_{\text{diag}} K_R+k_1 M^l_{\text{diag}}\right),
\end{align}
and, together with the lepton doublets (in the above formalism), are also encoded into the Yukawa interactions of Eq.\eqref{eq.yukawa}.

The implementation of the $\Delta_R$-type lepton-Higgs interactions is again done in the same way. In particular, the expression for the $h_M$ matrix in Eq.\eqref{eq.hm} becomes
\begin{align}
h_M=\frac{1}{\sqrt{2}v_R}\,K_R^T M^\nu_{\text{diag}} K_R.
\label{eq.hmascribing}
\end{align}
This definition, however, applies only to the lepton-$\Delta_R$ term, which consists of right handed mixing matrices, and \textit{not} to the lepton-$\Delta_L$ Yukawa term. The reason for this is that in contrast to the above described $h_L$ and $\tilde{h}_L$ matrices, using eq.\eqref{eq.hmascribing} in the lepton-$\Delta_L$ Yukawa term $\overline{\strut{(L_{iL})}^c}\;\Sigma_L {(h_M)}_{ij}L_{jL}$ leads to products between left handed and right handed mixings. These products are slightly different in the above formalism, in which the lepton mixing is ascribed to the neutrino mixing, from the results of sec.\ref{sub.yukawa}, as we shall now demonstrate. The products are between the right handed $K_R$ matrices (appearing in both sides of the $h_M$ expression, see Eq.\eqref{eq.hmascribing}) and the left handed mixings of the lepton doublets. One product includes the left handed neutrino mixing $K_R \cdot K_L^\dagger$, and the other includes the left handed charged lepton mixing $K_R\cdot\mathbb{1}$. In the first product the quasi-manifest matrix (see the definition of the diagonal matrix $W^l$ in Eq.\eqref{eq.quasi-lepton}) is present ($ K_RK^\dagger_L={V_R^\nu}^\dagger V^l_R{V_L^l}^\dagger V^\nu_L={V_R^\nu}^\dagger W^l V^\nu_L$) in contrast to this product in sec.\ref{sub.yukawa} (which is ${V_R^\nu}^\dagger V^\nu_L$); in the second product the quasi-manifest matrix is absent (Eq.\eqref{eq.quasi-lepton} defines $V^l_L$ in terms of $V^l_R$) - again in contrast to the same product in sec.\ref{sub.yukawa} (which is ${V_R^\nu}^\dagger V^l_L=K_RW^l$)\footnote{The discrepancy in the products between the current section formalism and sec.\ref{sub.yukawa} does not change upon replacing $L \leftrightarrow R$ in Eq.\eqref{eq.quasi-lepton}.}. The model file therefore uses the following alternative definition of the $h_M$ matrix for the lepton-$\Delta_L$ type terms, which incorporates the quasi-manifest relation in the charged-lepton sector:
\begin{align}
h_M=\frac{1}{\sqrt{2}v_R}\,W^l K_R^T M^\nu_{\text{diag}} K_R W^l.
\label{eq.hmascribing2}
\end{align}
As a result of the above reasoning, the proper Yukawa matrices for the lepton-$\Delta_R$ term (Eq.\eqref{eq.hmascribing}) and lepton-$\Delta_L$ term (Eq.\eqref{eq.hmascribing2}) are encoded, respectively, as
\begin{verbatim}
yHM1[a_,b_] -> 1/(vR*Sqrt[2]) Mr[a,b]
\end{verbatim}
and
\begin{verbatim}
yHM2[a_,b_] -> Module[{sp1,sp2},
	       1/(vR*Sqrt[2]) Wl[a,sp1] Mr[sp1,sp2] Wl[sp2,b]]
\end{verbatim}
where \verb+vR+ denotes the Higgs VEV $v_R$, the matrix \verb+Mr+ is the implemented $M_R$ matrix of Eq.\eqref{eq.useful2} rederived in terms of $K_R$, and \verb+Wl+ denotes the $W^l$ matrix. These encoded definitions yield the correct lepton-$\Delta_{L,R}$ Yukawa terms (for example, the definition of \verb+yHM2+ returns the result of Eq.\eqref{eq.Yukawa-lepton-final2} for the left handed singly charged Higgs Yukawa terms discussed above).

\subsection{Gauge boson and Higgs eigenstates}
The model file also defines the gauge eigenstates of the gauge and the Higgs bosons in terms of mass eigenstates. The gauge boson eigenstates are given in Eqs.\eqref{chargedbosons} and \eqref{neutralbosons} and are implemented accordingly. For example, the eigenstates of $\vec{W}_L$ (denoted as \verb+Wi+) are implemented as follows:
\begin{verbatim}
Wi[mu_,1] -> ((Wbar[mu]*cxi+W2bar[mu]*sxi)+(W[mu]*cxi+W2[mu]*sxi))
             /Sqrt[2],
Wi[mu_,2] -> ((Wbar[mu]*cxi+W2bar[mu]*sxi)-(W[mu]*cxi+W2[mu]*sxi))
             /(I*Sqrt[2]),
Wi[mu_,3] -> cw*cphi*Z[mu]+cw*sphi*Z2[mu]+sw*A[mu]
\end{verbatim}
where \verb+W+, \verb+W2+, \verb+Z2+, and \verb+A+ are the gauge boson mass eigenstates, multiplied by the proper mixing angles.

The Higgs gauge eigenstates are extracted as function of the mass eigenstates by diagonalizing the Higgs mass matrix as discussed in sec. \ref{sec.potential}. The precise form of these eigenstates is given in \cite{gluza}. The model file directly defines these eigenstates based on the constraints given in sec. \ref{sec.potential}.

\subsection{Parameter control by the user}
\label{partameter-control}
The model file allows the user to control certain parameters and thus adjust masses, mixings and interaction strength. These parameters include coupling constants, fermion masses, mixing matrix elements, Higgs VEVs and parameters in the Higgs potential, and are given in Table \ref{tab:adjustable} of \ref{App:AppendixA}.

The quark mixing matrices $U^{CKM}_L$ and  $U^{CKM}_R$ are related via Eq.\eqref{eq.ckm}. This relation is written in the model file as
\begin{verbatim}
Value -> {CKMR[a_,b_] -> WU[a,a]*CKML[a,b]*WD[b,b]}
\end{verbatim}
where the three matrices \verb+CKML+, \verb+WU+ and \verb+WD+ can be set by the user. In particular, the user can adjust the \verb+CKML+ matrix by setting the values of its elements via the external parameters \verb+s12+, \verb+s13+ and \verb+s23+  (see \ref{App:AppendixB}),
whereas setting the \verb+CKMR+ matrix elements is done in a non-direct manner following the above definition;
 setting \verb+WU[i,i]+ and \verb+WD[j,j]+ to $+1$ for every $i,j$ leads to the MLRSM where $U^{CKM}_R=U^{CKM}_L$,
 while setting at least one diagonal element of \verb+WU+ or \verb+WD+ to be negative leads to the QMLRS, where $\left(U^{CKM}_L\right)_{ij}=\pm \left(U^{CKM}_R\right)_{ij}$.

The dependency of the $K_L$ and $K_R$ matrices is looser than the $U^{CKM}$ matrices, and they are set and adjusted as follows.

The heavy-light mixing coefficients in Eq.\eqref{charged-current-majorana}, namely $\frac{M_D}{M_R}$ in the left handed current and $-\frac{M_D}{M_R}$ in the right handed current, are manifested in the three-generational $K_L$ and $K_R$, respectively (and explicitly in $K_{L\,\scriptsize{i+3,i}}$ and $K_{R\,\scriptsize{i,i}}$ where $i=1,2,3$). For simplicity, generation mixing is neglected and, in addition, the relation $M_D/M_R=\sqrt{M_{\text{light neutrino}}/M_{\text{heavy neutrino}}}$ was initially set to resemble the "vanilla" see-saw case \cite{vanilla}. The user can change the initial setting of these mixing coefficients, denoted as $V_e$, $V_\mu$ and $V_\tau$ (see \ref{App:AppendixA} and \ref{App:AppendixB}), by re-adjusting them.

The upper block of $K_L$ ($K_{L\,\scriptsize{i,j}}$ $i,j=1,2,3$), which connects the charged leptons with the light neutrinos, corresponds to the PMNS matrix of the SM \cite{PMNS1}, and can be adjusted via the parameters \verb+sL12+, \verb+sL13+ and \verb+sL23+ (see \ref{App:AppendixB}, \ref{App:AppendixB}) which define a unitary PMNS. The PMNS matrix, however, may deviate from unitarity as a result of the heavy-light neutrino mixing of the MLRSM, thus necessitating the independent adjustment of chosen matrix-elements (for further information and non-unitary fits see for example \cite{PMNS2}). 
The user can therefore choose, instead of adjusting the relevant external parameters mentioned above, to re-define each desired matrix-element of the $K_L$ and $K_R$ matrices in the \verb+FeynRules+ model file itself, and then re-translate it to the matrix-element generator. This second option, despite being graceless, is a preferable alternative to inserting a large number of external parameters into the model file and by thus making it too cumbersome and slowly processed.

The lepton sector also contains, analogically to the quark sector, a QMLRSM control matrix, namely $W^l$, whose diagonal elements can be set by the user in a similar way to the setting of $W^u$ and $W^d$ (see discussion in sec \ref{sub.yukawa}).

The user can control the coupling constants as well. This is done through the interaction strengths which can be set by the external parameters \verb+aEWM1+ and \verb+aS+ of the electroweak and strong interactions, respectively.

Controlling the Higgs mass eigenvalues is performed through the parameters in the Higgs potential (\verb+lambda[1..4]+, \verb+rho[1..4]+ and \verb+alpha[1..3]+ - the zero valued $\beta_i$ parameters are absent from the model file - see discussion in sec \ref{sec.potential}) and through the Higgs VEVs. Expressions of model parameters such as coupling constants, Higgs masses, gauge boson masses, mixing angles and potential parameters, all adjustable by the user through the parameters in Table \ref{tab:adjustable}, can be found in \ref{App:AppendixB}.

Controlling other parameters in the file (other than those in Table \ref{tab:adjustable}) is possible, but consistency should be kept. For example, the user can choose to switch the external parameter \verb+vR+ (the right Higgs triplet VEV) with the gauge boson mass \verb+MW2+. He should then set \verb+vR+ as a function of the new external parameters, e.g. \verb+MW+ and \verb+MW2+: \verb#vR=MW2/MW*Sqrt[(k1^2+k2^2)/2]# (where the approximation $v_R\gg k_1,\,k_2$ is used).

\section{Validation and output data}
\label{sec.Validation}
The model Implementation was validated closely following the requirements given in the manual of \cite{feynrules}. These include:
\begin{enumerate}
\item
The Feynman rules for the model file were calculated in \verb+Mathematica+ using the command
\begin{verbatim}
FeynmanRules[LLR]
\end{verbatim}
and were then matched with the Lagrangian terms in \cite{gluza}. A complete scan of the Lagrangian terms described in this work was performed, followed by a verification of some representative MLRSM process results from the model file by comparison with the literature.
\item
SM cross-sections calculated using the model file were compared and matched with their known values. In particular, values of the relevant parameters were chosen to reproduce the SM case. In total, more than 100 processes are tested, and the results (given with a 4 digits precision) agree well in all cases.
\item
A total of more than 250 processes which were based on the MLRSM model file were tested on matrix element generators and on the \verb+FeynRules+ program automatic $1\rightarrow2$ calculation function. The results from the different programs were found to match, either accurately (for $1\rightarrow2$ decays) or statistically (for Monte-Carlo evaluations of $2\rightarrow2$ scattering cross sections).
\item
High energy unitarity cancellations were tested and demonstrated.
\end{enumerate}
\subsection{Comparison with literature}
Table \ref{tab.comparison} compares the results calculated by \verb+FeynRules+ for the model file with 
the corresponding expressions given in the literature (\cite{classic} and \cite{gluza}). The Lagrangian vertices are compared and verified by explicitly identifying the relevant Feynman rule.
This Table is divided into green headlines of one or more Lagrangian terms followed by verified vertices extracted from 
these terms. As the manifest/quasi manifest LRSM Lagrangian contains a large number of vertices, only selected ones are accompanied by implicit Feynman rules. For each selected vertex, a list of other verified vertices of the same type is given underneath, below the dashed line.

\begin{figure}[htbp!]
\includegraphics[scale=0.66]{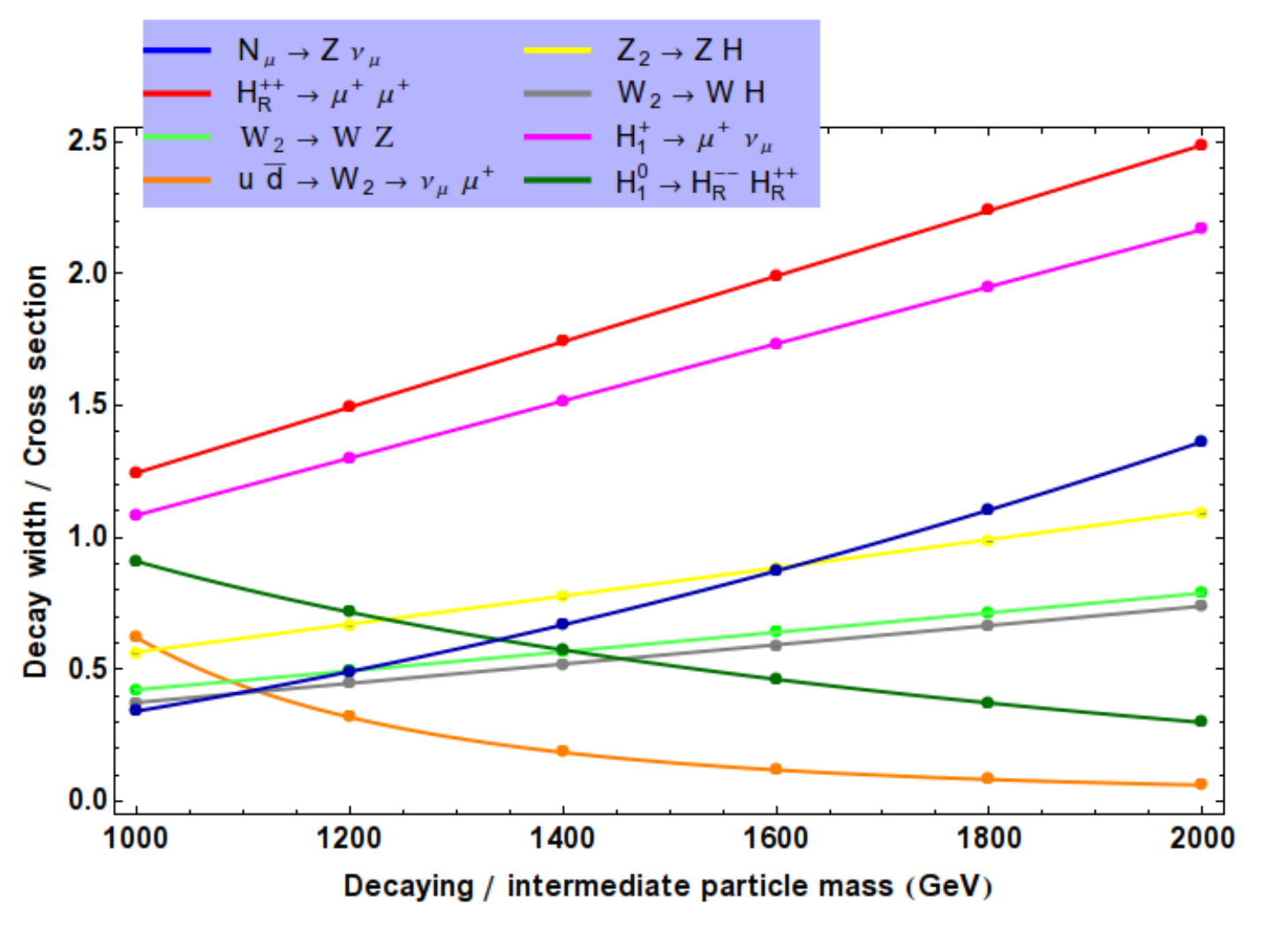}
\caption{Cross-section and decay widths of processes from Table \ref{tab.comparison2}. The dots correspond to values calculated numerically by the model file  while the curves correspond to the theoretical results. The measuring unit is $\text{GeV}$ for the decays, except from $N_\mu \rightarrow Z \, \nu_\mu$ which is given in units of $10^{-8}\,\text{GeV}$. The cross-section of $u \, \bar{d} \, \rightarrow W_2 \, \rightarrow \mu^+ \, \nu_\mu$ is given in units of $10^{-8}\,\text{pb}$.}
\label{TestPlot2-figure}
\end{figure}

In addition to the verification of the model file Lagrangian, some results produced by the model file were compared with 
the corresponding literature based results. The comparison was made for a range of processes and parameter values, for which the output of the model file was cross checked with independent calculations and with results from the literature. Table \ref{tab.comparison2} lists verified cross-sections and widths of some characteristic MLRSM processes which are either explicitly shown in the literature or calculated using literature-based guidelines given for similar processes. For each process (tested by the model file) presented in the left column a corresponding verified expression (i.e. a cross-section or a decay width in agreement with the model file result) is shown in the middle column, and the relevant verification source appears in the right column. Some representative results from Table \ref{tab.comparison2} are shown in Fig.~\ref{TestPlot2-figure}.

\subsection{Comparison with the SM}
\label{subsec.MLRSM-SM}
The processes chosen for the comparison with the SM were tested by taking the appropriate
SM limit of the parameters of the MLRSM model file and confronting them 
with results computed using the default \verb+FeynRules+ SM model file which can be found for reference in the \verb+FeynRules+ model database. In particular, the SM limit of the MLRSM model file was obtained by the following adjustments:
\begin{enumerate}
\item
Elimination of the gauge boson mixings by
\begin{itemize}
\item
setting $v_R\rightarrow \infty$, i.e. a very large number, which eliminates with neutral gauge boson mixing angle ($\phi=0$), and also leads to $M_{W_2,Z_2}\rightarrow \infty$.
\item
setting $k_2=0$, which leads to $W-W_2$ decoupling ($\xi=0$)\footnote{The expression for $k_2$ in the model file is derived from the expression of $W$ in the MLRSM \eqref{eq.W-bosons}.},
\end{itemize}
\item
Assigning values to the parameters of the Higgs potential according to the following guidelines (see the expressions for the Higgs masses in \ref{App:AppendixB}):
\begin{itemize}
\item
$\lambda_1=\lambda_\text{SM}$ (the $H$ Higgs particle is assumed accordingly to be the SM Higgs),
\item
$ 0 < 2\rho_1 < \rho_3 $,
\item
$ 0<\alpha_3,\rho_2 $.
\end{itemize}
\item
Assigning the following values to the Majorana neutrino masses:
\begin{align}
M_{\text{Light neutrinos}}=0,\;M_{\text{Heavy neutrinos}}\rightarrow\infty,
\end{align}
which leads to decoupling of the light and heavy neutrinos.
\end{enumerate}
 
The programs used were \verb+FeynRules+ (using its inner automatic $1\rightarrow2$ calculation function \cite{inner}) for the $1\rightarrow2$ decays and \verb+CalcHEP+ for the $2\rightarrow2$ processes. Total decays of the SM gauge bosons, Higgs and top quark are 
compared in Table \ref{tab.1to2sm}. 
The comparison of $2\rightarrow2$ processes is given in Table \ref{tab.2to2sm}. 
This Table is divided into different combinations of particle types (i.e. fermion, scalar, boson). 
Note that the cross sections/decay widths in the Tables \ref{tab.1to2sm} and \ref{tab.2to2sm} 
have to be multiplied by the exponent factor given in the column denoted by \textit{Exp.}.

\subsection{Comparing MLRSM results in matrix element generators}
\label{subsec.MLRSM-ch-mg}
Processes based on the MLRSM model file which were tested on different programs are shown in 
Table \ref{tab.1to2fr} ($1\rightarrow2$ decays) and in Table \ref{tab:ch-mg.comp} ($2\rightarrow2$ processes). 
The $1\rightarrow2$ decays were tested on \verb+CalcHEP+, \verb+MadGraph5_aMC@NLO+ and \verb+FeynRules+ (using its inner automatic $1\rightarrow2$ calculation function), and the $2\rightarrow2$ processes were tested on \verb+CalcHEP+ and \verb+MadGraph5_aMC@NLO+. The external (user controlled) parameter values used to obtain the cross sections (see \ref{appendix.comparsion-settings}) were chosen somewhat arbitrarily, by only requiring that
\begin{itemize}
\item The mass of $H$ (the SM Higgs) be in proximity to its known value,
\item The heavy gauge bosons (produced by the Higgs VEV $v_R$) be observable at the LHC, i.e. of a mass scale of $\sim1-\unit[10]{TeV}$.
\end{itemize}
The Monte Carlo results for the $2\rightarrow2$ processes in Table \ref{tab:ch-mg.comp} are divided into different combinations of particle types. These results are given with a three digit precision (in contrast to the other tables) as higher precision would give no additional information. Unlike Table \ref{tab.2to2sm}, in Table \ref{tab:ch-mg.comp} we do not present explicitly the Monte-Carlo uncertainties. The reason is that the \verb+FeynRules+ web interface \cite{feynrulesweb}, which we used for evaluating a large number of cross sections, often returned uncertainties of different scale for the \verb+CalcHEP+ and \verb+MadGraph5_aMC@NLO+ calculations. Therefore, in 
Table \ref{tab:ch-mg.comp} we demonstrate the statistical consistency of the results from the two programs by using the value of $\chi^2$, given as \cite{feynrules}
\begin{align}
\chi^2=\sum_{\text{i=CH,MG}}\left(\frac{\sigma_i-\sigma_b}{\Delta\sigma_i}\right)^2
\label{chi-square}
\end{align}
where $\sigma_i$ is the cross section produced by the relevant generator in the unitary gauge ($i=$\verb+CalcHEP+ ($\text{CH}$), \verb+MadGraph5_aMC@NLO+ ($\text{MG}$)), $\Delta\sigma_i$ is the Monte Carlo uncertainty returned by that generator and $\sigma_b$ is the best value for the cross section, defined as
\begin{align}
\sigma_b=\frac{\sum\limits_i\sigma_i/(\Delta\sigma_i)^2}{(\sum\limits_{\footnotesize{\text{i}}}1/\Delta\sigma_i^2)}.
\end{align}
Considering a theoretical $\chi_\mathcal{THEORY}^2$ distribution with one degree of freedom\footnote{The number of degrees of freedom is, for the current comparison, one, as two data files with unitary gauge are compared, see e.g., \cite{feynrules}.} and a standard deviation of $\sigma_\mathcal{THEORY}$ (with the $\chi^2$ distribution values: $1\sigma_\mathcal{THEORY}=1.00,\,2\sigma_\mathcal{THEORY}=4.00,\,3\sigma_\mathcal{THEORY}=9.00$ \cite{statistics}), and examining the complete list shown in Table \ref{tab:ch-mg.comp} of $2\rightarrow2$ processes tested in \verb+CalcHEP+ and \verb+MadGraph5_aMC@NLO+, we find that out of 200 processes
\begin{itemize}
\item[] The $\chi^2$ of 140 processes ($70\%$) are distributed within $1\sigma_\mathcal{THEORY}$ range,
\item[] The $\chi^2$ of 192 processes ($96\%$) are distributed within $2\sigma_\mathcal{THEORY}$ range and
\item[] The $\chi^2$ of 198 processes ($99\%$) are distributed within $3\sigma_\mathcal{THEORY}$ range.
\end{itemize}
Thus, the distribution of the $\chi^2$ defined in Eq.\eqref{chi-square} corresponds well to the normal distribution of the theoretical $\chi_\mathcal{THEORY}^2$, indicating a correct implementation of the model file \cite{feynrules}.
\subsection{Cancellations of unitarity violating contributions}
\label{subsec.cancellations}
\begin{figure}[h]
\hspace{-0.8cm}
\includegraphics[scale=0.47]{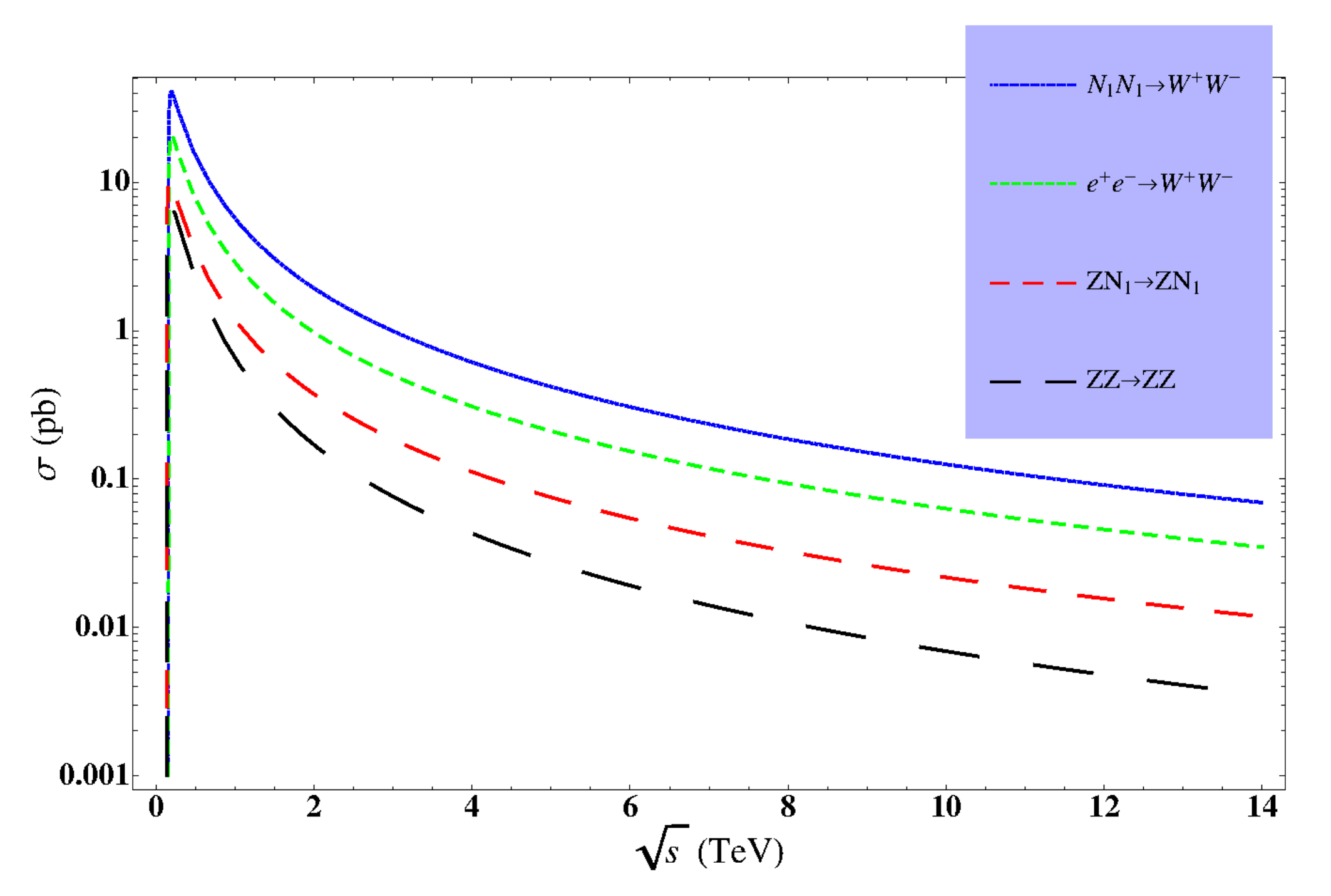}
\caption{Lowest order total cross-sections to MLRSM processes with MLRSM-SM common external particles which manifest unitarity restoration.}
\label{unitarity-plot}
\end{figure}

\begin{figure}[h]
\hspace{-0.8cm}
\includegraphics[scale=0.47]{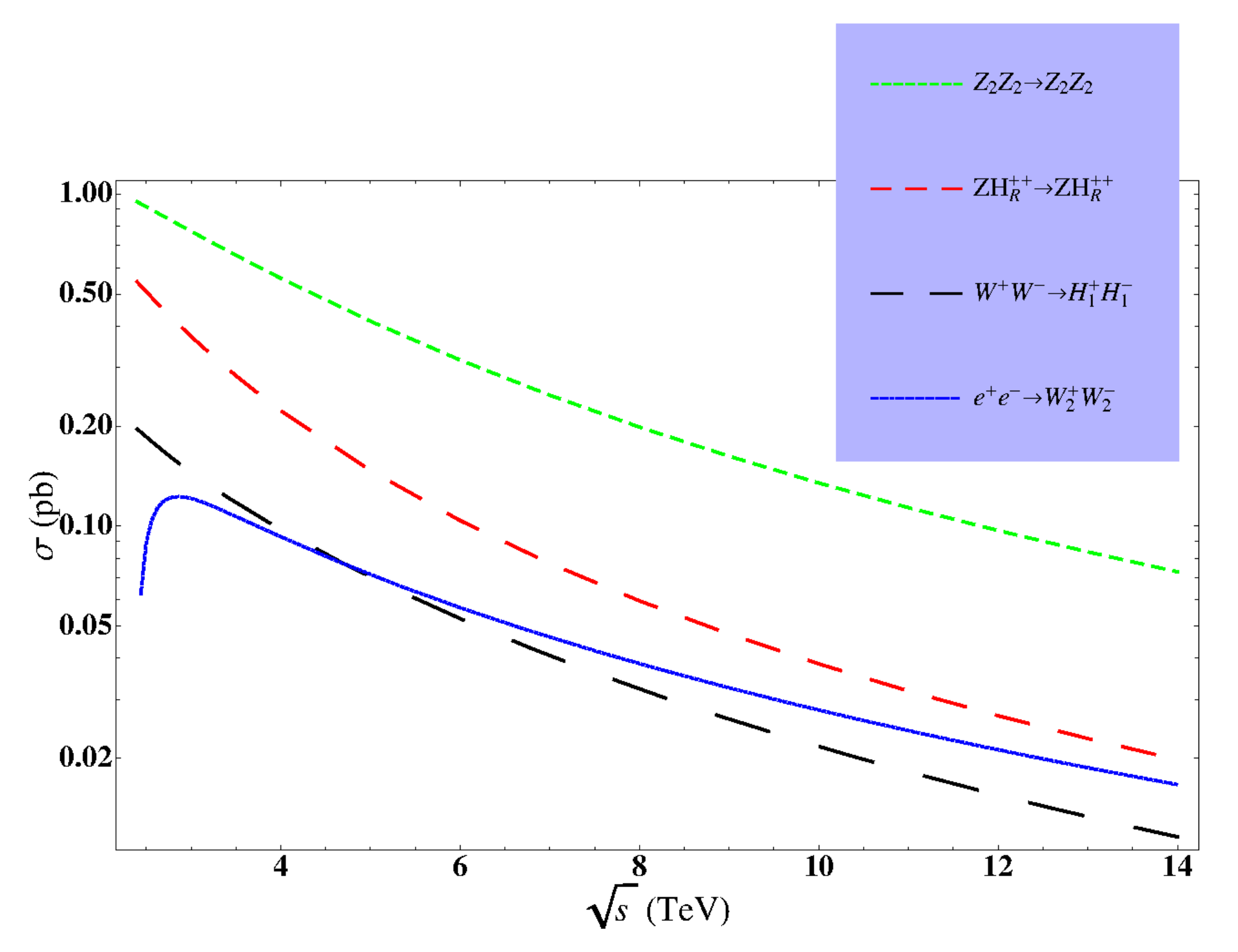}
\caption{Lowest order total cross-sections which manifest unitarity restoration to MLRSM processes containing some beyond SM particles.}
\label{unitarity-plot2}
\end{figure}

Tables \ref{tab.2to2sm} and \ref{tab:ch-mg.comp} contain a number of processes with contributions which violate unitarity at high energies. However, as required in renormalizable models, these contributions cancel each other, thus allowing the corresponding cross sections to preserve unitarity. Examples of processes which manifest the restoration of unitarity at high energies are presented in figure \ref{unitarity-plot} (for MLRSM processes with MLRSM-SM common external particles) and in figure \ref{unitarity-plot2} (for MLRSM processes containing some beyond-SM external particles).
\newpage
\newpage
\newpage
\begin{footnotesize}
 \renewcommand{\arraystretch}{2.1}
\begin{longtable}[c]{| l |}
\caption{Comparison of the LRSM model file results with the corresponding literature expressions (Refs. \cite{classic} and \cite{gluza})}
\label{tab.comparison} \\
\hline
\endfirsthead
\multicolumn{1}{c}%
{{\bfseries \tablename\ \thetable{} -- continued from previous page}} \\
\hline\hline
\endhead
\multicolumn{1}{|r|}{{Continued on next page}} \\ \hline
\endfoot
\endlastfoot
\hline
\rowcolor{green!20}Lagrangian term: $ \sum_{\Psi=(Q),(L)}\bar{\Psi}_L\gamma^\mu(i\partial_\mu+g\frac{\tau}{2}\vec{W}_{L\mu}+g'\frac{Y}{2} B_\mu )\Psi_L +(L\rightarrow R)$\\*
\hline
$\bar{u}W^{+\mu}d \Rightarrow$ $\frac{g}{\sqrt{2}}\gamma^\mu(P_L\cos\xi\,U^{CKM}_L-P_R\sin\xi\,U^{CKM}_R) $ \\*
\hdashline
$\bar{u} W_2^+ d$ and the vertices h.c. \\
\hline
$\; \bar{u} Z^\mu u \Rightarrow$ $\frac{ig}{2}\cos\phi\cos\Theta_w \gamma_\mu P_L + \frac{ig'\sin\phi\sin\Theta_w\gamma_\mu P_L}{6\cos\Theta_w}-\frac{ig\cos\phi\sin^2\Theta_w\gamma_\mu P_L}{6\cos\Theta_w}+ \frac{ig'\sin\phi\sin\Theta_w \gamma_\mu P_R}{6\cos\Theta_w}$  \\*
$- \frac{2ig\cos\phi\sin^2\Theta_w\gamma_\mu P_R}{3\cos_w}-\frac{ig\sin\phi\sqrt{1-2\sin^2\Theta_w}\gamma_\mu P_R}{2\cos_w} $\\*
\hdashline
$\bar{d}Z^\mu d,\;\bar{u}A^\mu u,\;\bar{d}A^\mu d,\;\bar{u}Z_2^\mu u,\bar{d}Z_2^\mu d.$ \\
\hline
$\bar{N}W^{+\mu}l \Rightarrow \frac{ig}{\sqrt{2}}\gamma^\mu(P_L\cos\xi\,K_L-P_R\sin\xi\,K_R)$ \\*
\hdashline
$\bar{N} W_2^+ l$ and the vertices h.c. \\
\hline

\end{longtable}
\renewcommand{\arraystretch}{2}
\begin{longtable}{| l | l | l |}
\caption{Comparison with literature based calculations / results.}
\label{tab.comparison2}\\
    \hline\hline
\rowcolor[gray]{0.85}\rule{0pt}{4ex}Process & Verified result (formula) & Source  \\
    \hline
\endfirsthead
\multicolumn{3}{c}%
{{\bfseries \tablename\ \thetable{} -- continued from previous page}} \\
\hline\hline
\rowcolor[gray]{0.85}\rule{0pt}{4ex}Process & Verified result (formula) & Source  \\ \hline
\endhead
\hline
\multicolumn{3}{|r|}{{Continued on next page}} \\ \hline
\endfoot
\endlastfoot
$N_4 \stackrel{\mathrm{W_{1,2},Z}}{\longrightarrow} N_1 e^+ e^-$ & $\frac{G_F^2 m_{N_4}^5}{192\pi^3}(\sin^2{\xi}+\Delta K_{L41}^2),\;\Delta=\frac{1}{4}-\sin^2\Theta_W+2\sin^4\Theta_W$ & Ref.\cite{mohapatra} \\
\hline
$N_4 \stackrel{\mathrm{Z}}{\longrightarrow} N_1 N_1 N_1$ & $\frac{G_F^2 m_{N_4}^5}{192\pi^3}K_{L41}^2$ & Ref.\cite{mohapatra} \\
\hline
$N_4 \stackrel{\mathrm{W_2}}{\longrightarrow} e^+ + \text{hadrons}$ & $\frac{9\,G_F^2 m_{N_4}^5}{192\pi^3}\left(\frac{M_{W_1}}{M_{W_2}}\right)^4$ (diagonal CKM) & Ref.\cite{mohapatra} \\
\hline
$\mu^+ \stackrel{\mathrm{W_2}}{\longrightarrow} N_2 e^+ N_1$ & $\frac{G_F^2 m_\mu^5}{192\pi^3}\left(\frac{M_{W_1}}{M_{W_2}}\right)^4I\left(\frac{m_e^2}{m_\mu^2}\right),\, I(x)=1-8x+8x^3-x^4-12x^2ln(x)$ & Ref.\cite{collider} \\
\hline
$c \stackrel{\mathrm{W_2}}{\longrightarrow} s(e^+N_1$ &  $\frac{5G_F^2 m_\mu^5}{192\pi^3}\left(\frac{M_{W_1}}{M_{W_2}}\right)^4I\left(\frac{m_s^2}{m_c^2}\right)$ & Ref.\cite{collider} \\
$\hspace{1cm}  / \mu^+N_2/ u\bar{d})$ & & \\
\hline
$e^+ e^- \stackrel{\mathrm{Z,Z_2}}{\longrightarrow} \mu^+ \mu^-$ & $\dfrac{g^4m_{Z,Z_2}^4}{12\pi m_W^4}(g_V^2+g_A^2)\dfrac{s}{(s-m_Z^2)^2+(\Gamma_{Z,Z_2}m_{Z,Z_2})^2}$ & Ref.\cite{commins} \\
& $g_V$: The vec. coupling (coef. of $\gamma_\mu$) of vertex $\bar{l}lZ$ (or $Z_2$) & \\
\hline
\end{longtable}
\end{footnotesize}
\clearpage
\begin{longtable}{| l | l | l | l |}
\caption{Comparison of FeynRules results for total $1\rightarrow2$ decay widths modes given within the SM and within the MLRSM 
taken in the SM limit, see also text.}
\label{tab.1to2sm}\\
    \hline\hline
\rowcolor[gray]{0.85}\rule{0pt}{4ex}Decaying particle & $\Gamma_{\text{SM}}$ (GeV) & $\Gamma_{\text{MLRSM}}$ (GeV) & Exp. \\
    \hline
\endfirsthead
\multicolumn{4}{c}%
{{\bfseries \tablename\ \thetable{} -- continued from previous page}} \\
\hline
\rowcolor[gray]{0.85}\rule{0pt}{4ex}Decaying particle & $\Gamma_{\text{SM}}$ (GeV) & $\Gamma_{\text{MLRSM}}$ (GeV) & Exponent \\ \hline
\endhead
\hline
\multicolumn{4}{|r|}{{Continued on next page}} \\ \hline
\endfoot
\endlastfoot
\rule{0pt}{4ex}
    \hspace {0.9cm} $W$ & 2.1172 & 2.1172 & \\
    \hspace {1cm} $Z$ & 2.5253 & 2.5253 & \\
    \hspace {1cm} $H$ & 5.8092 & 5.8092 & $10^{-3}$ \\
    \hspace {1cm} $t$ & 1.5124 & 1.5124 & \\
    \hline
\end{longtable}
\begin{footnotesize}
\renewcommand{\arraystretch}{1.3}
\begin{longtable}{ l  l  l  l  l  l }
\caption{Cross sections, calculated with CalcHEP, of $2\rightarrow2$ processes in the SM and in the SM limit of 
the MLRSM model file, see also text.}
\label{tab.2to2sm} \\
\hline \hline
\rowcolor[gray]{0.85}\rule{0pt}{4ex}MLRSM Process & $\sqrt{s}$ & $\text{PT}_\text{cut}$ & $\sigma_{\text{SM}}(pb)$ & $\sigma_{\text{MLRSM}}(pb)$ & Exp. \\
\hline
 \endfirsthead
\multicolumn{6}{c}%
{{\bfseries \tablename\ \thetable{} -- continued from previous page}} \\
\hline
\rowcolor[gray]{0.85}\rule{0pt}{4ex}MLRSM Process & $\sqrt{s}$ & $\text{PT}_\text{cut}$ & $\sigma_{\text{SM}}(pb)$ & $\sigma_{\text{MLRSM}}(pb)$ & Exp. \\ \hline
\endhead
\hline
\multicolumn{6}{r}{{Continued on next page}} \\ \hline
\endfoot
\endlastfoot
\hline
\rule{0pt}{4ex}
\hspace{-0.1cm}$ e^-	,	e^-	\rightarrow	e^-	,	e^- 	$ & $ 200.0 $ & $ 50.0 $ & $ 1.2459\pm0.0001 $ & $ 1.2459 \pm 0.0001$ & $10^2$\\
$ e^-,e^+\rightarrow	u,\bar{u} $ & $ 200.0 $ & $ 50.0 $ & $ 4.1375\pm0.0003 $ & $ 4.1377\pm0.0003 $ & \\
$ e^-,e^+\rightarrow	t,\bar{t} $ & $ 1376.0 $ & $ 344.0 $ & $ 7.7181\pm0.0005 $ & $ 7.7180\pm0.0005 $ & $10^1$ \\
$ e^-, \mu^-	\rightarrow e^- , \mu^-	$ & $ 7000.0 	$ & $ 0.0 $ & $ 1.3491 \pm 0.0001$ & $ 1.3491 \pm 0.0001$  & $10^1$\\
$ \mu^-,	\mu^-\rightarrow	\mu^-,\mu^- $ & $ 200.0 $ & $ 50.0 $ & $1.2567\pm0.0001 $ & $1.2567\pm0.0001$ & $10^{-2}$\\	
$ \mu^-, \mu^+ \rightarrow N_1, N_1$\footnote{\label{note1}The processes in Table \ref{tab.2to2sm} with a final state of two MLRSM Majorana neutrinos agree, for the settings applied in \ref{subsec.MLRSM-SM}, with the analogous SM processes with Dirac neutrino and anti-neutrino replacing the Majorana pair (this can be seen by applying the Feynman rules for the MLRSM vertex $ZN_iN_j\,(i,j=1..3)$ given in Table \ref{tab.comparison}), in accordance with refs.\cite{kayser} and \cite{collider}.} & $ 200.0 	$ & $ 0.0 $ & $ 	1.0486 \pm 0.0001 $ & $ 	1.0486 \pm 0.0001$ & \\
$ \mu^-,	\mu^+\rightarrow	u,\bar{u} $ & $ 200.0 $ & $ 50.0 $ & $	4.3506\pm0.0003 $ & $ 4.3506\pm0.0003 $ & \\	
$ \mu^-,	\mu^+\rightarrow	c,\bar{c} $ & $ 200.0 $ & $ 50.0 $ & $	4.3501\pm0.0003 $ & $ 4.3502\pm0.0003 $ & \\
\hline
\end{longtable}
\end{footnotesize}
\begin{footnotesize}
\renewcommand{\arraystretch}{1.3}
\begin{longtable}{| l | l | l | l | l | l |}
\caption{Total decay widths in $1\rightarrow2$ decay mode for the MLRSM (FeynRules vs. CalcHEP vs. MadGraph5). The chosen MLRSM parameter settings and the resulting particle masses can be found in \ref{appendix.comparsion-settings}.}
\label{tab.1to2fr}\\
\hline \hline
\rowcolor[gray]{0.85}Decaying particle & Mass (GeV) & $\Gamma_{\text{FR}}$ (GeV) & $\Gamma_{\text{CH}}$ (GeV) & $\Gamma_{\text{MG}}$ (GeV) & Exp. \\
\hline
 \endfirsthead
\multicolumn{6}{c}%
{{\bfseries \tablename\ \thetable{} -- continued from previous page}} \\
\hline
\rowcolor[gray]{0.85}Decaying particle & Mass (GeV) & $\Gamma_{\text{FR}}$ (GeV) & $\Gamma_{\text{CH}}$ (GeV) & $\Gamma_{\text{MG}}$ (GeV) & Exp. \\
\hline
\endhead
\hline
\multicolumn{5}{|r|}{{Continued on next page}} \\
\hline
\endfoot
\endlastfoot
\hline
\rule{0pt}{4ex}
    \hspace {0.9cm} $W$ & $8.0399*10^1$ & 2.1172 & 2.1170 & 2.1172 & \\
    \hspace {1cm} $Z$ & $9.1188*10^1$ & 2.5241 & 2.5240 & 2.5241 &\\
    \hspace {0.9cm} $W_2$ & $1.1975*10^3$ & 4.2190 & 4.2190 & 4.2190 & $10^1$\\
    \hspace {1cm} $Z_2$ & $2.0012*10^3$ & 5.8952 & 5.8950 & 5.8952 & $10^1$\\
    \hspace {1cm} $H$ & $1.2000*10^2$ & 5.8141 & 5.8140 & 5.8141 & $10^{-3}$ \\    
    \hspace {1cm} $H^0_1$ & $1.5021*10^3$ & 3.0498 & 3.0500 & 3.0498 & $10^2$ \\
    \hspace {1cm} $H^0_2$ & $1.9700*10^2$ & 7.8144 & 7.8140 & 7.8144 & $10^{-9}$ \\
    \hspace {1cm} $H^0_3$ & $2.0057*10^3$ & 1.8508 & 1.8510 & 1.8508 & $10^{-1}$\\
\hline
\end{longtable}
\end{footnotesize}
\begin{footnotesize}
\renewcommand{\arraystretch}{1.3}
\begin{longtable}{|| p{4.0cm} || p{1.5cm} || p{1.5cm} | p{1.4cm} | p{1.4cm} | p{1.0cm} | p{1.6cm} |}
\caption{Cross sections of $2\rightarrow2$ processes in the MLRSM (CalcHEP vs. MadGraph5). The chosen MLRSM parameter settings and the resulting particle masses can be found in \ref{appendix.comparsion-settings}.  Although the guidelines for the validation of implemented models indicate that the particle decay widths should be set to zero when comparing different matrix element generators, we retained these parameters in 30 processes, denoted by an asterisk. This is done in order to enhance the sensitivity to probe possible discrepancies between the programs due to interferences among gauge bosons and Higgs particles.}
\label{tab.2to2comp-ch-mg}\\
\hline \hline
\rowcolor[gray]{0.85}MLRSM Process & $\sqrt{s}$ & $\text{PT}_\text{cut}$ & $\sigma_{\text{CH}}(pb)$ & $\sigma_{\text{MG}}(pb)$ & Exp. & $\chi^2$ \\
\hline
 \endfirsthead
\multicolumn{7}{c}%
{{\bfseries \tablename\ \thetable{} -- continued from previous page}} \\
\hline
\rowcolor[gray]{0.85}MLRSM Process & $\sqrt{s}$ & $\text{PT}_\text{cut}$ & $\sigma_{\text{CH}}(pb)$ & $\sigma_{\text{MG}}(pb)$ & Exp. & $\chi^2$\\ \hline
\endhead
\hline
\multicolumn{7}{|r|}{{Continued on next page}} \\ \hline
\endfoot
\endlastfoot
    \hline
   \rule{0pt}{4ex}
   \hspace{-0.2cm} $ e^-, e^-\rightarrow e^-,e^- $ & $2100.0 $ & $	100.0 $ & $	3.288 $ & $ 3.286 $ & $ 10^1 $ & $9.3*10^{-1}$ \\
$ e^-, e^+\rightarrow e^-, e^+ $ (*)& $	2100.0 $ & $	200.0 $ & $6.858 $ & $ 6.861 $ & $  $ & $ 1.7*10^{-1}$ \\
$ e^-, e+	 \rightarrow N_1 N_1$ (*) &	$ 2100.0 $ &	$ 0.0 $ & $	5.727 $ & $	5.728 $ & $10^1$ & $1.8*10^{-2} $ \\
$e^-	, e^+	\rightarrow	N_4	, N_4 $ & $3200.0$ &	$500.0 $ & $	1.589 $ & $	1.587 $ & $10^{-1}$ & $	1.2*10^{-3} $ \\
$e^-	, e^+	\rightarrow	d , \bar{d} $ &	$ 2100.0 $ &	$ 100.0 $ &  $9.081$ & $	9.084 $ & $10^{-1} $ & $7.4*10^{-2} $ \\
$e^-	, \mu^+	\rightarrow	N_1	, N_2 $ & $2100.0 $ & $0.0 $ &  $5.742 $ & $	5.739 $ & $10^1$ & $1.6 $ \\
$e^-	, \mu^+	\rightarrow	N_1	, N_5 $ & $2100.0 $ & $200.0 $ & $1.432 $ & $1.424 $ & $10^{-4}$ & $	1.2 $ \\
$e^-	, \tau^+\rightarrow	N_1	, N_3 	$ &	$2100.0 $ &	$0.0 $ &  $	5.739 $ & $	5.740 $ & $10^1$ & $	3.9 $ \\
\end{longtable}
\end{footnotesize}
\FloatBarrier
\section{Summary}
\label{sec.Conclusion}
We described in this paper the manifest/quasi manifest left-right symmetric model and its implementation in \verb+FeynRules+. The model file constructs a model Lagrangian in which one can define the model parameters solely in physical basis (in contrast to having to define Yukawa coupling matrices directly). The model is composed of Majorana type neutrinos, and features a simple and direct control of its quasi-manifest relation matrices. We performed a wide range of validation tests for the model, completely scanning its Lagrangian while comparing it to the literature. In light of the appealing new features that the left-right symmetric models offer (to address some of the shortcomings of the standard model), the implementation presented in this work can serve as an important tool for the search for new physics at the LHC era.










\appendix
\clearpage
\setcounter{table}{0}
\section{Notation and adjustable (external) parameters}
\label{App:AppendixA}
The MLRSM input parameters and the corresponding symbols in the model file are collected in Table \ref{tab:parameters}
(following the notation in \cite{gluza}), while the list of the adjustable external parameters in the model file 
is given in Table \ref{tab:adjustable}.
\begin{footnotesize}
\renewcommand{\arraystretch}{1.17}
\begin{longtable}{|p{4.2cm}|l|l|}
\caption{The MLRSM parameter names and the corresponding symbols in the FeynRules model file}
\label{tab:parameters}\\
\hline \hline
\rowcolor[gray]{0.85}Category & LRSM symbol & \verb+Feynrules+ model symbol \\
\hline
 \endfirsthead
\multicolumn{3}{c}%
{{\bfseries \tablename\ \thetable{} -- continued from previous page}} \\
\hline\hline
\rowcolor[gray]{0.85}Category & LRSM symbol & \verb+Feynrules+ model symbol \\ \hline
\endhead
\hline
\multicolumn{3}{|r|}{{Continued on next page}} \\ \hline
\endfoot
\endlastfoot
    \hline
\rule{0pt}{4ex}
\multirow{3}{*}[3mm]{\vbox{Fermion doublets (Gauge eigenstates)}} &  Quark doublets: $Q_{iL}$, $Q_{iR}$ &  \verb+QL+,  \verb+QR+ \\[-0.4cm]
 & Lepton doublets: $L_{iL}$, $L_{iR},$ & \verb+LL+, \verb+LR+ \\[0.1cm]
& Charge Conj. (Lepton doublets): & \\
& $(L^c_i)_L$, $(L^c_i)_R\,\,(i=1,2,3)$ & \verb+LCL+, \verb+LCR+ \\
\hline
\rule{0pt}{4ex}
\multirow{2}{*}[3mm]{Gauge boson fields} & $W_{Li}$, $W_{Ri}$, $B$ $\,\,(i=1,2,3)$ & \verb+Wi+, \verb+WRi+, \verb+B+\\
(Gauge eigenstates) & & \\
\hline
\rule{0pt}{4ex}
    \multirow{18}{4cm}[3mm]{\vbox{Particles names (physical states)}} & SM Gauge bosons: $W$, $Z$, $A$, $g$ & \verb+W+, \verb+Z+, \verb+A+, \verb+G+,  \\
    & Extra SM Gauge bosons: $W_2$, $Z_2$ & \verb+W2+, \verb+Z2+ \\
    & Up type quarks: $u$, $c$, $t$ & \verb+u+, \verb+c+, \verb+t+ (class name: \verb+uq+) \\
    & Down type quarks: $d$, $s$, $b$ & \verb+d+, \verb+s+, \verb+b+ (class name: \verb+dq+) \\
    & Charged leptons: $e$, $\mu$, $\tau$ & \verb+e+, \verb+mu+, \verb+ta+ (class name: \verb+l+) \\
    \cline{2-3}
    & Light neutrinos: $N_{1(e)}$, $N_{2(\mu)}$, $N_{3(\tau)}$ & \verb+NeL+, \verb+NmL+, \verb+NtL+ \\
    & Heavy neutrinos: $N_{4(e)}$, $N_{5(\mu)}$, $N_{6(\tau)}$ &  \verb+NeH+, \verb+NmH+, \verb+NtH+ \\
    & & (class name: \verb+Nl+) \\
     \cline{2-3}
    & Neutral Higgs scalars: $H^0_0$, $H^0_1$,  & \verb+H+, \verb+H01+, \verb+H02+, \verb+H03+ \\
	& $H^0_2$, $H^0_3$ & \\
    & $A^0_1$, $A^0_2$ (Neutral Higgs & \verb+A01+, \verb+A02+ \\
    & pseudoscalars) & \\
    & $H^\pm_1$, $H^\pm_2$, $\delta^{\pm\pm}_L$, $\delta^{\pm\pm}_R$ (Charged & \verb+HP1+, \verb+HP2+, \verb+HPPL+, \verb+HPPR+ \\
    	& Higgs scalars) & \\
    	& $\tilde{G}^0_1$, $\tilde{G}^0_2$ (Neutral Goldstone & \verb+G01+, \verb+G02+  \\
    	& bosons)\footnote{The model file uses the unitary gauge, so that all the
    Goldstone modes are omitted in the Feynman rules calculation.}& \\
    	& $G^\pm_L$, $G^\pm_R$ (Charged Goldstone & \verb+GPL+, \verb+GPR+  \\
    	& bosons)\footnotemark[\value{footnote}] & \\
\hline
\rule{0pt}{4ex}
Particle masses & $M_{\text{Relevant Particle}}$ & The letter M + Particle name\\
\hline
\rule{0pt}{4ex}
\multirow{2}{*}{Decay widths} &  \multirow{2}{*}[3mm]{$\Gamma_{\text{Relevant Particle}}$} & Either zero or \\ & & the letter W + Particle name \\
\hline
\rule{0pt}{4ex}
Mixing Matrices & $U_L^{CKM}$, $U_R^{CKM}$ (Quark) & \verb+CKML+, \verb+CKMR+,  \\
\rule{0pt}{4ex}
Mixing Matrices&$K_L$, $K_R$ (Lepton)& \verb+KL+, \verb+KR+ \\
\rule{0pt}{4ex}
Mixing Matrix&$U$ (Neutral Higgs Scalars)& \verb+U+ \\
\hline
\rule{0pt}{4ex}
\multirow{5}{*}{Mixing parameters} & $s_{12},\;s_{23},\;s_{33}$ ($U_L^{CKM}$) & \verb+s12+, \verb+s23+, \verb+s23+  \\
& $c_{12},\;c_{23},\;c_{33}$ ($U_L^{CKM}$) & \verb+c12+, \verb+c23+, \verb+c23+  \\
& $V_e,\; V_\mu,\; V_\tau$ ($K_{L,R}$ h-l. mixing)& \verb+VKe+, \verb+VKmu+, \verb+VKta+ \\
& $sL_{12},\;sL_{23},\;sL_{33}$ ($K_L$) & \verb+sL12+, \verb+sL23+, \verb+sL23+  \\
& $cL_{12},\;cL_{23},\;cL_{33}$ ($K_L$) & \verb+cL12+, \verb+cL23+, \verb+cL23+  \\
\hline
\rule{0pt}{4ex}
Quasi manifest matrices & $W_l$, $W_u$, $W_d$ & \verb+Wl+, \verb+WU+, \verb+WD+ \\
\hline
\rule{0pt}{4ex}
Quasi manifest parameters & $W_{laa}$, $W_{uaa}$, $W_{daa},\;(a=1..3)$ & \verb+Wl11-33+, \verb+WU11-33+, \verb+WD11-33+ \\
\hline
\rule{0pt}{4ex}
\multirow{3}{4cm}{Mixing angles} & $\sin\Theta_W$, $\cos\Theta_W$ (Weinberg) & \verb+sw+, \verb+cw+, \\
& $\sin\xi$, $\cos\xi$, (Charged gauge boson)  & \verb+sxi+, \verb+cxi+,  \\
& $\sin\phi$, $\cos\phi$. (Neutral gauge boson) & \verb+sphi+, \verb+cphi+ \\
\hline
\rule{0pt}{4ex}
\multirow{2}{3.5cm}{Higgs VEVs} & $k_1$, $k_2$, $k_+$, $k_-$, $v_L$, $v_R$ & \verb+k1+, \verb+k2+, \verb+vev+, \verb+kminus+, \verb+vL+, \verb+vR+ \\
&sign($k_2$) & \verb+sk2+ \\
\hline
\rule{0pt}{4ex}
Higgs multiplets & $\phi$ , $\tilde{\phi}$, $\Delta_{L,R}$ & \verb+BD+, \verb+BDtilde+, \verb+LT+, \verb+RT+ \\
\hline
\rule{0pt}{4ex}
\multirow{2}{3.5cm}{Higgs multiplet field components} & $\phi^0_{1,2}$, $\phi^\pm_{1,2}$ & \verb+Phi01+, \verb+Phi02+, \verb+PhiP1+, \verb+PhiP2+ \\
& $\delta_{L,R}^0$, $\delta_{L,R}^\pm$  & \verb+H0L+, \verb+H0R+, \verb+HPL+, \verb+HPR+ \\
& $\delta_{L,R}^{\pm\pm}$ & \verb+HDPL+, \verb+HDPR+ \\
\hline
\rule{0pt}{4ex}
\multirow{2}{3cm}{Parameters in the Potential} & $\mu_{1..3}^2$, $\lambda_{1..4}$, & \verb+musq[1]+..\verb+[3]+, \verb+lambda1+..\verb+4+ \\ & $\rho_{1..4}$, $\alpha_{1..3}$ & \verb+rho1+..\verb+4+, \verb+alpha1+..\verb+3+ \\
\hline
\rule{0pt}{4ex}
\multirow{2}{*}{Yukawa matrices} & $h_Q$ $\tilde{h}_Q$, $h_L$, $\tilde{h}_L$, & \verb+yQ+, \verb+yQtilde+, \verb+yL+, \verb+yLtilde+, \\ & $h_M$ & \verb+yHM1+, \verb+yHM2+ (both relate to $h_M$) \\
\hline
\rule{0pt}{4ex}
\multirow{2}{3cm}{Diagonal mass matrices} & $(M_{U})_\text{diag}$, $(M_{D})_\text{diag}$, & \verb+yMU+, \verb+yDO+ \\ &   $(M_{l})_\text{diag}$, $(M_{\nu})_\text{diag}$ & yML, yNL \\
\hline
\rule{0pt}{4ex}
\multirow{2}{*}{couplings}  & $\alpha(M_Z)$, $\alpha_s(M_Z)$, $G_f$ & \verb+aEW+ \verb+aS+, \verb+Gf+ \\
  & $e$, $g$, $g'$, $g_s$ & \verb+ee+, \verb+gw+, \verb+g1+, \verb+gs+ \\
\hline
\end{longtable}
\renewcommand{\arraystretch}{1.17}
\begin{longtable}{|l|l|l|}
\caption{The external (user controlled) parameters in the model file.}
\label{tab:adjustable}\\
\hline \hline
\rowcolor[gray]{0.85} LRSM parameter & \verb+FEYNRULES+ model symbol & Section in the file \\
\hline
 \endfirsthead
\multicolumn{3}{c}%
{{\bfseries \tablename\ \thetable{} -- continued from previous page}} \\
\hline
\rowcolor[gray]{0.85} LRSM parameter & \verb+FEYNRULES+ model symbol & Section in the file \\ \hline
\endhead
\hline
\multicolumn{3}{|r|}{{Continued on next page}} \\ \hline
\endfoot
\endlastfoot
    \hline
\rule{0pt}{4ex}
    \multirow{6}{*}[3mm]{Fermion masses} & \verb+MU+,\verb+MC+,\verb+MT+ & \multirow{5}{*}{\bigcell{c}{\hspace{-1.6cm} Particle classes / \\ \hspace{-0.2cm} Fermions: physical fields}} \\
    & \verb+MD+, \verb+MS+, \verb+MB+ & \\
    & \verb+Me+, \verb+Mmu+, \verb+Mta+, & \\
    & \verb+MNeL+, \verb+MMmL+, \verb+MNtL+ & \\
    & \verb+MNeH+, \verb+MMmH+, \verb+MNtH+ & \\
    \hline
    \rule{0pt}{4ex}
     \multirow{2}{*}[3mm]{Gauge boson masses} & \verb+MW+, \verb+MZ+ & Particle classes / \\
     & & Gauge bosons: physical vector fields \\
    \hline
\rule{0pt}{4ex}
Decay widths & \verb+M+ (in mass symbol) $\rightarrow$ \verb+W+ & Relevant particle definition\\
\hline
\rule{0pt}{4ex}
Higgs VEVs & \verb+k1+, \verb+vR+, \verb+sk2+ & Parameters / External Parameters \\
\hline
 \multirow{3}{*}{\bigcell{c}{\hspace{0mm} Pararameters in the \\ \hspace{-1.65cm} Potential}} & \verb+lambda1+..\verb+4+,  & Parameters / External Parameters \\
 & \verb+rho1+ .. \verb+4+, & Parameters / External Parameters \\
 & \verb+alpha1+ .. \verb+3+ & Parameters / External Parameters \\
\hline
\rule{0pt}{4ex}
\multirow{3}{*}{Mixing parameters} & \verb+s12+, \verb+s13+, \verb+s23+ & Parameters / External Parameters \\
& \verb+VKe+, \verb+VKmu+, \verb+VKta+ & Parameters / External Parameters \\
& \verb+sL12+, \verb+sL13+, \verb+sL23+ & Parameters / External Parameters  \\
\hline
\rule{0pt}{4ex}
\multirow{3}{*}{Q.Manifest parameters} & \verb+WU11+, \verb+WU22+, \verb+WU33+ & Parameters / External Parameters \\
 & \verb+WD11+, \verb+WD22+, \verb+WD33+ & Parameters / External Parameters \\
 & \verb+Wl11+, \verb+Wl22+, \verb+Wl33+ & Parameters / External Parameters \\
\hline
\rule{0pt}{4ex}
Constants & \verb+aEWM1+, \verb+Gf+, \verb+aS+ & Parameters / External Parameters \\
\hline
\end{longtable}
\end{footnotesize}
\section{Parameters, masses and mixing angles in the MLRSM} \label{App:AppendixB}
The external parameters in \ref{App:AppendixA} constitute the internal model ingredients, which are based on the Lagrangian of Ref.\cite{gluza}.\footnote{The expressions for $\mu_i^2$ also appear in refs.\cite{classic,barenboim}. In addition, the quark mixing parameters are shown in Ref.\cite{pdg} and the lepton mixing parameters are shown in Refs.\cite{kayser,PMNS1,PMNS2}.}  These internal parameters are indirectly controlled by the user via the expressions given below (the $H^0_0$, $H^0_1$ and $H^0_2$ masses are given in the approximation $v_R \gg k_+$):
\begin{align}
& k_2=sign(k_2)\,*\,\sqrt{k_+^2-k_1^2},\quad\text{(where}\;k_+=\unit[246]{GeV}\;\text{is set as the SM Higgs VEV)} \nonumber \\
& k_{-}^2=k_1^2 - k_2^2,\;\; \cos\Theta_W=\frac{M_W}{M_Z}, \;\; \tan{2\xi}=-\frac{2k_1k_2}{v_R^2}, \;\; \sin{2\phi}=-\frac{g^2k_+^2\sqrt{cos{2\Theta_W}}}{2\cos^2\Theta_W(M_{Z_2}^2-M_Z^2)}, \nonumber \\
& g_s=\sqrt{4\pi\alpha_s(M_Z)}, \quad e=\sqrt{4\pi\alpha(M_Z)}, \quad g=\frac{e}{\sin\Theta_W}, \quad g'=\frac{e}{\sqrt{\cos{2\Theta_W}}}, \nonumber \\
& M^2_{W_2}=\frac{g^2}{4}\Big\{k_+^2+v_R^2+\sqrt{v_R^4+4\,k_1^2\,k_2^2}\Big\}, \nonumber \\
& M^2_{Z_2}=\frac{1}{4}\Big\{g^2k_+^2+2v_R^2(g^2+g'^2)+\sqrt{{\big[g^2\,k_+^2+
2v_R^2(g^2+g'^2)\big]}^2-4g^2(g^2+2g'^2)\,k_+^2\,v_R^2 }\Big\}, \nonumber \\ 
& M^2_{H^0_0}\approx 2\,k_+^2\,\left(\lambda_1+\frac{4k_1^2k_2^2}{k_+^4}(2\lambda_1+\lambda_3)+2\lambda_4\frac{2k_1k_2}{k_+^2}\right), \nonumber \\
& M^2_{H^0_1}\approx \frac{1}{2}\,{\alpha_3\,v_R^2}\,\frac{k_+^2}{k_-^2}, \quad M^2_{H^0_2} \approx 2\,\rho_1\,v_R^2, \quad M^2_{H^0_3}=\frac{1}{2}v_R^2\,(\rho_3-2\rho_1), \nonumber \\
& M^2_{A^0_1}=\frac{\alpha_3\,v_R^2}{2}\,\frac{k_+^2}{k_-^2}-2k_+^2(2\lambda_2-\lambda_3), \quad M^2_{A^0_2}=\frac{1}{2}v_R^2(\rho_3-2\rho_1), \nonumber \\
& M^2_{H^{\pm}_1}=\frac{1}{4}(\alpha_3\,(k_-^2))+\frac{1}{2}v_R^2(\rho_3-2\rho_1), \quad M^2_{H^{\pm}_2}=\frac{1}{4}\alpha_3\left(k_-^2+2\frac{k_+^2}{k_-^2}\,v_R^2\right), \nonumber \\
& M^2_{\delta^{\pm\pm}_L}=\frac{1}{2}\left(\alpha_3\,(k_-^2)+v_R^2(\rho_3-2\rho_1)\right), \quad M^2_{\delta^{\pm\pm}_R}=\frac{1}{2}\left(\alpha_3\,(k_-^2)+4v_R^2\rho_2\right), \nonumber \\
& \mu^2_1= v_R^2 \left(\frac{\alpha_1}{2}-\frac{\alpha_3 k_2^2}{2\,(k1^2-k2^2)}\right)+(k_+^2\lambda_1+2k_1k_2\lambda_4), \nonumber \\
& \mu^2_2= v_R^2\left(\frac{\alpha_2}{2}+\frac{\alpha_3 k_1 k_2}{4\,(k_-^2)}\right)+k_1\,k_2(2\lambda_2+\lambda_3)+\frac{\lambda_4k_+^2}{2}, \nonumber \\
& \mu_3^2=\rho_1\,v_R^2+\frac{\alpha_1\,k_+^2}{2}+2\alpha_2\,k_1\,k_2+\frac{\alpha_3\,k_2^2}{2}, \nonumber \\
& \text{(the Higgs potential parameters $\mu_i$, $\lambda_i$, $\rho_i$ and $\alpha_i$ are defined in Eq.\eqref{eq.potential}).} \nonumber \\
\footnotesize
& U_L^{CKM} =\left(\begin{array}{ccc}
c_{12}c_{13} & s_{12}c_{13} & s_{13} \\
-s_{12}c_{23}-c_{12}s_{23}s_{13} & c_{12}c_{23}-s_{12}s_{23}s_{13} & s_{23}c_{13} \\
s_{12}s_{23}-c_{12}c_{23}s_{13} & -c_{12}s_{23}-s_{12}c_{23}s_{13} & c_{23}c_{13} \end{array} \right), \nonumber \\[8pt]
& c_{12}=\sqrt{1-s_{12}^2},\quad c_{13}=\sqrt{1-s_{13}^2}, \quad c_{23}=\sqrt{1-s_{23}^2} \nonumber \\[5pt]
& W_{u/d/l}=\left(\begin{array}{ccc}
W_{u11}/W_{d11}/W_{l11} & 0 & 0 \\
0 & W_{u22}/W_{d22}/W_{l22} & 0 \\
0 & 0 & W_{u33}/W_{d33}/W_{l33} \end{array} \right), \nonumber \\[12pt]
& U_R^{CKM}=W^U U_L^{CKM} W^D, \nonumber \\[12pt]
& K_L=\left(\begin{array}{ccc}
cL_{12}\,cL_{13} & sL_{12}\,cL_{13} & sL_{13} \\
-sL_{12}\,cL_{23}-c_{12}\,sL_{23}\,sL_{13} & cL_{12}\,cL_{23}-sL_{12}\,sL_{23}\,sL_{13} & sL_{23}\,cL_{13} \\
sL_{12}\,sL_{23}-cL_{12}\,cL_{23}\,sL_{13} & -cL_{12}\,sL_{23}-sL_{12}\,cL_{23}\,sL_{13} & cL_{23}\,cL_{13} \nonumber \\[8pt]
V_e & 0 & 0 \\
0 & V_\mu & 0 \\
0 & 0 & V_\tau
\end{array} \right), \nonumber \\[12pt]
& cL_{12}=\sqrt{1-{sL_{12}}^2},\quad cL_{13}=\sqrt{1-{sL_{13}}^2}, \quad cL_{23}=\sqrt{1-{sL_{23}}^2} \nonumber \\[5pt]
& K_R=\left(\begin{array}{ccc}
-V_e & 0 & 0 \\
0 & -V_\mu & 0 \\
0 & 0 & -V_\tau \\
1 & 0 & 0 \\
0 & 1 & 0 \\
0 & 0 & 1 \\
\end{array} \right) \nonumber \\[12pt]
\end{align}
\section{Higgs physical eigenstates}
\label{appendix.Higgs physical eigenstates}
Obtaining the fields eigensystem is done by diagonalizing the squared-mass matrix given in Eq.\eqref{eq.mass-matrix}. The diagonalization is simplified by the constraints in sec. \ref{sec.potential}. The eigenstates consist of\footnote{We use the unitary gauge}
\begin{enumerate}
\item Four neutral scalar eigenstates $H$, $H^0_1$, $H^0_2$, $H^0_3$,
\item Two neutral pseudoscalar eigenstates $A^0_1$, $A^0_2$,
\item Four singly charged scalar eigenstates $H^\pm_1$, $H^\pm_2$, 
\item Two doubly charged scalar eigenstates $H^{\pm\pm}_L$, $H^{\pm\pm}_R$.
\end{enumerate}
The corresponding eigenvalues/masses are given in \ref{App:AppendixB}. The non-physical Higgs fields can then be written in terms of the above eigenstates as follows:
\begin{align}
& \phi^0_1=\frac{1}{\sqrt{2}}\left(k_1 + U_{11} H + U_{12} H^0_1 + U_{13} H^0_2 +i\frac{k_2}{k_+}\,A^0_1\right), \nonumber \\
& \phi^0_2=\frac{1}{\sqrt{2}}\left(k_2 + U_{21} H + U_{22} H^0_1 + U_{23} H^0_2+i\frac{k_1}{k_+}\,A^0_1\right), \nonumber \\
& \delta_L^0=\frac{1}{\sqrt{2}}\left(v_L+H^0_3+i\,A^0_2\right), \nonumber \\
& \delta_R^0=\frac{1}{\sqrt{2}}\left(v_R+U_{31}H+U_{32}H^0_1+U_{33}H^0_2\right), \nonumber \\
&\phi^{\pm}_1=\frac{k_1}{k_+\,\sqrt{1+{(\frac{k_-^2}{\sqrt{2}k_+v_R})}^2}}\,H_2^{\pm} \nonumber \\
&\phi^{\pm}_1=\frac{k_2}{k_+\,\sqrt{1+{(\frac{k_-^2}{\sqrt{2}k_+v_R})}^2}}\,H_2^{\pm} \nonumber \\
&\delta_L^{\pm}=H_1^{\pm}, \nonumber \\[4pt]
&\delta_R^{\pm}= \frac{1}{\sqrt{1+{(\frac{\sqrt{2}k_+v_R}{k_-^2})}^2}}\,H_2^{\pm} \nonumber \\[4pt]
&\delta_{L,R}^{\pm\pm}=H_{L,R}^{\pm\pm},
\end{align}
where $U$ is an orthogonal mixing matrix which diagonalizes the (real) squared-mass matrix of the neutral Higgs scalars\footnote{The $U$ mixing matrix appears in the model file containing the supplement $\text{\_ }$\texttt{mix} in its name. The other model versions use the simplifying approximation $v_R \gg k_1,k_2$ which lead to the $\phi^0_1$, $\phi^0_2$ and $\delta^0_R$ states given in \cite{gluza}.}. Therefore
\begin{align}
\left(\begin{array}{c}
\phi^{0{\cal{R}}}_1 \\
\phi^{0{\cal{R}}}_2 \\
\delta^{0{\cal{R}}}_R
\end{array} \right)=
\left(\begin{array}{ccc}
U_{11} & U_{12} & U_{13} \\
U_{21} & U_{22} & U_{23} \\
U_{31} & U_{32} & U_{33} \\
\end{array} \right)
\left(\begin{array}{c}
H \\
H^0_1 \\
H^0_2
\end{array} \right).
\end{align}
The matrix elements of $U$ can be calculated numerically by loading the model in a \verb+Mathematica+ session. Upon using the commands\\

\verb+ComputeMassMatrix[poten2,Mix->"1s"]+

\verb+MassMatrix["1s"]+\\[0.5cm]
one recovers the scalar (squared mass) block in the neutral Higgs sector, namely
\begin{align*}
\hspace{-7cm}{\cal{M}}^2=
\end{align*}
\small
\begin{align}
& \left(\begin{smallmatrix}
2k_1^2\lambda_1\!+\!4k_1k_2\lambda_4\!+\!k_2^2\left(\frac{\alpha_3vR^2}{2k_-^2}+4\lambda_2+\lambda_3\right) & 2\lambda_4(k_1^2\!+\!k_2^2)\!+\!\frac{1}{2}k_1k_2\left(4\lambda_1\!+\!8\lambda_2\!+\!4\lambda_3\!-\!\frac{\alpha_3v_R^2}{k_-^2} \right) & v_R\,(\alpha_1 k_1\!+\!2\alpha_2k_2) \\
& & \\
2\lambda_4(k_1^2\!+\!k_2^2)\!+\!\frac{1}{2}k_1k_2\left(4\lambda_1\!+\!8\lambda_2\!+\!4\lambda_3\!-\!\frac{\alpha_3v_R^2}{k_-^2} \right) & 2k_1^2\,(2\lambda_2\!+\!\lambda_3)\!+\!4k_1k_2\lambda_4\!+\!\frac{1}{2}\alpha_3v_R^2\!+\!k_2^2\left(2\lambda_1\!+\!\frac{\alpha_3v_R^2}{2k_-^2}\right) & v_R\,(2\alpha_2k_1\!+\!k_2\,(\alpha_1\!+\!\alpha_3)) \\
& & \\
v_R\,(\alpha_1 k_1\!+\!2\alpha_2k_2) & v_R\,(2\alpha_2k_1\!+\!k_2\,(\alpha_1\!+\!\alpha_3)) & 2\rho_1 v_R^2 \\
\end{smallmatrix} \right). \nonumber
\end{align}
\normalsize
\vspace{-3mm}
\begin{equation}
\end{equation}
In order to get the numerical elements of $U$ and the masses of $H$, $H^0_1$ and $H^0_2$, one can use the \verb+ASperGe+ program \cite{asperge} by typing\\

\verb+WriteAsperge[poten2]+\\[0.5cm]
after assigning numerical values to all the parameters appearing in the mass matrix. 

\section{\mbox{Parameter settings and gauge boson masses for Tables \ref{tab.1to2fr} and \ref{tab:ch-mg.comp}}}
\label{appendix.comparsion-settings}
The parameter settings of the MLRSM model file for the process comparisons performed in Tables \ref{tab.1to2fr} and \ref{tab:ch-mg.comp} are as follows:
\begin{itemize}
\item
Masses, Widths (in GeV)
\begin{align}
& M_{t}=173,\;M_{c}=1.27,\;M_{u}=2.55\,\times\,10^{-3}, \nonumber \\
& M_{b}=4.7,\;M_{s}=0.101,\;M_{d}=5.04\,\times\,10^{-3}, \nonumber \\
& M_{e}=5.11\,\times\,10^{-4},\;M_{\mu}=0.106,\;M_{\tau}=1.777, \nonumber \\
& M_{N_1}=10^{-8},\;M_{N_2}=10^{-8},\;M_{N_3}=10^{-8}, \nonumber \\
& M_{N_4}=100,\;M_{N_5}=100,\;M_{N_6}=100, \nonumber \\
& M_W=80.399,\; M_Z=91.188, \nonumber \\
& W_{\text{All particles}}=0 \hspace{0.5cm}\text{(applies only to Table } \ref{tab.2to2comp-ch-mg}).
\end{align}
\item
Higgs VEVs (in GeV)
\begin{align}
& k_1=227.91,\; v_R=2543.2.
\end{align}
\item
Parameters in the Higgs potential
\begin{align}
& \lambda_1=0.118,\;\lambda_2=0.2,\;\lambda_3=-0.234,\;\lambda_4=0, \nonumber\\
& \rho_1=0.003,\;\rho_2=0.001,\;\rho_3=1.25,\;\rho_4=0.125, \nonumber\\
& \alpha_1=0.5,\; \alpha_2=0.5,\; \alpha_3=0.5.
\end{align}
\item
Couplings
\begin{align}
& G_f=\unit[1.166\times10^{-5}]{GeV^{-2}},\; \alpha_s(M_Z)=0.1184, \; \alpha(M_Z)=\frac{1}{127.9},
\end{align}
\item
Mixing parameters
\begin{align}
& s12=0.221,\;s23=0.041,\;s13=0.0035, \nonumber \\
& V_e=\sqrt{M_{N_1}/M_{N_4}},\; V_\mu=\sqrt{M_{N_2}/M_{N_5}},\; V_\tau=\sqrt{M_{N_3}/M_{N_6}}, \nonumber \\
& W_{uaa}=W_{daa}=W_{laa}=1\;(a=1..3).
\end{align}
\end{itemize}
The resulting physical boson masses are (in \unit{GeV})

\begin{footnotesize}
\begin{tabular}{l l l l }
\addlinespace
\addlinespace
\hspace{0.1cm} $M_{W}=8.0399*10^1$, & $M_{Z}=9.1188*10^1$, & $M_{W_2}=1.1975*10^3$, & $M_{Z_2}=2.0012*10^3$, \\
\addlinespace
\hspace{0.1cm} $M_{H}=1.2000*10^2$, & $M_{H^0_1}=1.5021*10^3$, & $M_{H^0_2}=1.9700*10^2$, & $M_{H^0_3}=2.0057*10^3$,\\
\addlinespace
\hspace{0.1cm} $M_{A^0_1}=1.4763*10^3$, & $M_{A^0_2}=2.0057*10^3$,  & & \\
\addlinespace
\hspace{0.1cm} $M_{H^\pm_1}=2.0071*10^3$, & $M_{H^\pm_2}=1.5039*10^3$,  & & \\
\addlinespace
\hspace{0.1cm} $M_{H^{\pm\pm}_L}=2.0084*10^3$, & $M_{H^{\pm\pm}_R}=1.5420*10^2$. & & \\
\end{tabular}
\end{footnotesize}
\end{document}